\documentclass[%
superscriptaddress,
reprint,
showpacs,
 amsmath,amssymb,
 aps,
prd,
showkeys
]{revtex4-1}

\usepackage{graphicx}
\usepackage{dcolumn}
\usepackage{bm}
\usepackage[bookmarks=true,bookmarksopen=true, bookmarksnumbered=true]{hyperref}
\usepackage[mathlines]{lineno}
\usepackage{amsfonts, amsmath, amssymb}
\usepackage{xcolor}
\usepackage{float}
\usepackage{siunitx}
\usepackage[list=true, labelformat=brace]{subcaption}

\begin{document}


\title{Multipole vectors of completely random microwave skies 
for $l \leq 50$}
\author{Marvin Pinkwart}
\email{m.pinkwart@jacobs-university.de}
 \affiliation{Fakult\"{a}t f\"{u}r Physik, Universit\"{a}t Bielefeld, Postfach 100131, 33501 Bielefeld,
Germany}
\affiliation{Department of Physics and Earth Sciences, Jacobs University Bremen, Campus Ring 1, 28759 Bremen, Germany}
 
\author{Dominik J. Schwarz}%
 \email{dschwarz@physik.uni-bielefeld.de}
 \affiliation{Fakult\"{a}t f\"{u}r Physik, Universit\"{a}t Bielefeld, Postfach 100131, 33501 Bielefeld,
Germany}

\date{\today}

\begin{abstract}
The statistical cosmological principle states that observables on the celestial sphere 
are sampled from a rotationally invariant distribution. Previously certain large scale 
anomalies which violate this principle have been found, for example 
an alignment of the lowest multipoles with the cosmic dipole direction.
In this work we continue the search for possible anomalies using multipole vectors which represent a convenient tool for this purpose. In order to study the statistical behavior of multipole vectors, we revisit several construction methods.

We investigate all four full-sky 
foreground-cleaned maps from the Planck 2015
release with respect to four meaningful physical directions using computationally cheap statistics 
that have a simple geometric interpretation. We find that the full-sky SEVEM map deviates from all the other cleaned
maps, as it shows a strong correlation with the Galactic Pole and
Galactic Center. The other three maps COMMANDER, NILC and SMICA show a consistent behavior. On the largest angular scales, $l \leq 5$, as well as on intermediate scales, $l=20,21,22,23,24$, all of them 
are unusually correlated with the cosmic dipole direction. These scales coincide with the scales on which the angular power spectrum deviates from the 
Planck 2015 best-fit $\Lambda$CDM model. In the range $2 \leq l \leq 50$ as a whole there is no
unusual behavior visible globally. We do not find abnormal intramultipole correlation, i.e.\ correlation 
of multipole vectors inside a given multipole without reference to any outer direction. 

\end{abstract}

\maketitle


\section{Introduction}

High-fidelity observations of the cosmic microwave background (CMB) at the largest angular scales 
became available with data from the 
Wilkinson Microwave Anisotropy Probe (WMAP) \cite{WMAP}. Previous full-sky analyses 
based on data from the Cosmic Background 
Explorer (COBE) suffered at the largest angular scales from their limited capacity (only three 
frequency bands) to reliably separate the various foreground components from the CMB. 
Once confidence on foreground separation techniques was built, the WMAP data offered the 
potential to address the statistical isotropy of the observed temperature anisotropies at 
large angular scales. The property of statistical isotropy is a fundamental assumption in the analysis of the CMB and for the estimation of cosmological parameters. It was noted that the quadrupole and octopole seem to be aligned with 
each other \cite{deOliviera2004,Schwarz2004} and with the CMB dipole \cite{Schwarz2004} 
and an unexpected hemispherical asymmetry was revealed \cite{Eriksen2004}. 
More signs of violation of statistical isotropy have been found in several 
publications \cite{Hansen2004,Helling,Schwarz2007,Eriksen2007, Schwarz2009,Kim2010, 
Schwarz2010,Bennett2011}. 

On the other hand, already the WMAP data suggested that deviations from 
Gaussianity and from the angular power 
spectrum predicted by the $\Lambda$CDM model are negligibly small 
\cite{KomatsuNonGaussianity, KomatsuWMAP}. The analysis of the data from the 
Planck satellite confirmed these findings \cite{Planck2015I,Planck2015XIII,Planck2015XVII}, but 
at the same time confirmed the existence of isotropy anomalies \cite{Planck2015XVI}.

The full Planck mission data allowed to construct four precise full-sky maps that use different 
cleaning algorithms to remove the influences of the Milky Way \cite{Planck2015IX, Planck2015X}.
That analysis increases our confidence that the aforementioned isotropy anomalies are not due 
to instrumental effects, mistakes in the analysis pipeline, or unaccounted foregrounds, which should have been revealed by the wide frequency coverage of Planck.
A recent review collects the up-to-date knowledge about these isotropy anomalies \cite{Schwarz2015}. 

Analyses of the cosmic background radiation are conveniently performed by means of 
spherical harmonic coefficients and angular power spectra, 
or correlation functions in angular space. When investigating the CMB with respect to 
possible breaking of statistical isotropy, a third tool has become popular, namely multipole vectors (MPVs) \cite{MultiCopi,KatzWeeks,Helling}. 
While spherical harmonic coefficients transform with Wigner's symbols under a rotation of the celestial sphere, MPVs transform like ordinary three-vectors, i.e.\ they rotate rigidly with the temperature fluctuations on the sphere which makes them a convenient choice for isotropy analysis.

In this article we review the common construction methods for multipole vectors as well as their theoretical statistical behavior and use the four foreground cleaned full-sky maps from the Planck 2015 data analysis 
to investigate the statistical isotropy of the CMB, especially alignments of multipole vectors within 
a multipole or with external directions. It should be noted that the multipole vector method can only test for statistical isotropy if Gaussianity of the temperature fluctuations is assumed to hold since multipole vector statistics are only sensitive to deviations from a completely random distribution.

In Sec.~\ref{Spectrum} we review the basic definitions and properties of CMB data analysis by means of the angular power spectrum and we describe our convention of statistical isotropy. Then, in Sec.~\ref{multipole vectors} we give an overview over three convenient extraction methods for MPVs. In Sec.~\ref{Stat_props} we review the derivation of the probability distribution of MPVs. In Sec.~\ref{data} we shortly describe the Planck data used in the analysis. Section ~\ref{Analysis} is dedicated to the introduction of the statistics that we use for the analysis. Section \ref{directions} introduces the four astrophysical directions used in the analysis to estimate sources of multipole anomalies. In Sec.~\ref{results} we present the results before discussing them in Sec.~\ref{discussion} and giving a short conclusion and outlook in Sec.~\ref{conclusion}.

\section{Angular power spectrum}
\label{Spectrum}

The relative fluctuations of CMB temperature, which live on the celestial sphere, 
are conveniently decomposed according to the irreducible representations of the 
group of three-dimensional spatial rotations $SO(3)$, namely the orthonormal set 
of spherical harmonic functions,
\begin{equation}
\label{SphDecomp}
\frac{\delta T}{T_0}(\mathbf{e}) = \sum_{l=1}^{\infty} \sum_{m=-l}^l a_{lm} Y_{lm}(\mathbf{e}),
\end{equation}
with the radial unit vector $\mathbf{e} = (\cos(\phi)\sin(\theta),\sin(\phi)\sin(\theta),\cos(\theta))$ 
pointing towards the direction of observation. The contribution from a given integer number $l$,
\begin{equation}
\label{fl_def}
f_l(\mathbf{e}) = \sum_m a_{lm}Y_{lm}(\mathbf{e}),
\end{equation}
is called
a multipole of order $l$, which describes features at typical angular scales of about $\alpha_l = \pi/l$. 
The coefficients $a_{lm}$ are called spherical harmonic coefficients. 
Thanks to the orthonormality of $\{Y_{lm}\}$, i.e., $\int Y_{lm}Y_{l'm'}^* = \delta_{ll'}\delta_{mm'}$, 
the $a_{lm}$ can be calculated from $\delta T/T_0$ via integration
\begin{equation}
\label{ALMcalc}
a_{lm} = \int \! \mathrm{d}^3\mathbf{e} \, \frac{\delta T (\mathbf{e})}{T_0} Y_{lm}^* (\mathbf{e}).
\end{equation}

Since temperature fluctuations 
are real, and $Y_{lm}^* = (-1)^m Y_{l,-m}$, the spherical harmonic coefficients obey
\begin{equation}
\label{almrelation}
a^*_{lm} = (-1)^ma_{l,-m}.
\end{equation}

The particular pattern of the CMB temperature fluctuations cannot be predicted. Instead,  
temperature fluctuations are modeled as a real, random field on the sphere, 
or equivalently we model the measured spherical harmonic coefficients as 
realizations of an ensemble of random variables, subject to condition (\ref{almrelation}).

A fundamental assumption regarding the temperature fluctuations is statistical isotropy, 
that means
\begin{align}
\label{isotropy}
\begin{split}
&\forall \, R \in SO(3)\, \forall \, \mathbf{e}_1,\ldots, \mathbf{e}_n \in \mathcal{S}^2\, \forall \, n \in \mathbb{N}:\\
& \left\langle \prod_{i=1}^n \frac{\delta T}{T_0}(R \mathbf{e}_i) \right\rangle = \left\langle \prod_{i=1}^n \frac{\delta T}{T_0}(\mathbf{e}_i) \right\rangle,
\end{split}
\end{align}
where $\langle . \rangle$ denotes the expectation value of random fields, respectively the 
ensemble average over all ``possible universes''. Correlation functions of temperature fluctuations 
at different directions should only depend on the angle between them, respectively on the scalar 
products $\mathbf{e}_i\cdot \mathbf{e}_j$. 

Here, we take the point of view of an observer in three-dimensional space that acts with an element 
of $SO(3)$ on the sky. There exists also the other convention that the three degrees of freedom of 
three-dimensional rotations are split into a translation on $\mathcal{S}^ 2$ -- and invariance under such 
a transformation would then be called homogeneity -- and a rotation around a point on $\mathcal{S}^2$.
The invariance of the latter would then be associated with isotropy. In our work all three symmetry
operations are viewed as rotations and thus we only speak about isotropy.  

One usually argues that the smallness of the CMB temperature fluctuations provides empirical 
evidence for statistical isotropy of the Universe and cosmological inflation provides an argument on why 
the observed patch of the Universe should be isotropic. But there are \textit{a priori} no other reasons and 
a detailed study of the observed deviations from isotropy in $\delta T/T_0$ might reveal that 
statistical isotropy could be violated, for example due to primordial anisotropies. 

A trivial consequence of the definition of $\delta T$ is the vanishing of the one-point function, 
$\langle  \delta T(\mathbf{e})/T_0 \rangle = 0$ or $\langle a_{lm} \rangle = 0$. 
Thus the first nontrivial and most interesting object is the 
angular two-point correlation or the angular power spectrum $C_l$, which for an isotropic 
ensemble is given by $\langle a_{lm}^* a_{l'm'}\rangle = C_l \delta_{ll'}\delta_{mm'}$. 

As a result of the initial Gaussianity after inflation and the following linear evolution, this random 
field is assumed to be Gaussian in the standard theory. That means that higher correlations cannot 
carry independent information. Thus in the standard model of cosmology all cosmological information 
is encoded in the angular power spectrum $C_l$. 

\section{Multipole vectors}
\label{multipole vectors}

MPVs represent a tool for investigating CMB anisotropies in a very natural 
manner. They behave like ordinary three-vectors under rotation and they do not distinguish a certain reference 
frame, unlike the spherical harmonics which incorporate a reference to a chosen $z$-axis in 
their definition. In the following, we review different mathematical approaches to the description of functions on the sphere via MPVs. This is necessary to 
provide the appropriate tools for an analytic study of the statistical distribution of MPVs on completely random skies. While the algebraic and tensorial approaches yield recursive relations for direct calculation of the MPVs from a given spherical harmonic decomposition, the coherent state approach gives the MPVs as roots of a complex polynomial. With the help of the latter one can  calculate analytically the joint probability density given a fixed multipole.

\subsection{Origin and Sylvester's theorem}
\label{origin}
MPVs date back to Maxwell who introduced them in \cite{Maxwell}, in the study of interactions between monopoles. A monopole creates an electric potential proportional to $1/r$. Maxwell argued that the potential of two opposite-signed monopoles can be written as a directional derivative of the monopole potential $D_{\mathbf{v}}(1/r)$, where $\mathbf{v}$ denotes the linking vector between the point charges. He continued to the case of $3,4,\ldots$ interacting monopoles and received a potential of the form $D_{\mathbf{v}_1}\ldots D_{\mathbf{v}_l}(1/r)$ if $l$ monopoles are involved. Later on it has been noticed that any real, harmonic and homogeneous polynomial on $\mathbb{R}^3$ can be represented in that form. 

Let $f:\mathbb{R}^3 \rightarrow \mathbb{R}$ be a real, harmonic and homogeneous polynomial of degree $l$ in the variables $x,y,z$. That means $\Delta_{\mathbb{R}^3}f = 0$, implying $\Delta_{\mathcal{S}^2 } f = -l(l+1)f$, and $f(\lambda x, \lambda y, \lambda z) = \lambda^l f(x,y,z)$. Any such polynomial defines a polynomial $\widetilde{f} = f|_{\mathcal{S}^2}:\mathcal{S}^2 \rightarrow \mathbb{R}$ on the sphere and vice versa. Maxwell's MPV representation states that there exist $l$ unique unit directions $\mathbf{v}_1,\ldots, \mathbf{v}_l$, such that $f$ takes on the following form
\begin{align}
\label{MaxwellRep}
\begin{split}
f(x,y,z) &= (\mathbf{v}_1\cdot \nabla)\ldots(\mathbf{v}_l\cdot \nabla)\frac{1}{r(x,y,z)} \\
\text{with } r(x,y,z) &= \sqrt{x^2+y^2+z^2}.
\end{split}
\end{align}
This statement is known as Sylvester's theorem \cite{Sylvester}.

The expression \eqref{MaxwellRep} is equivalent to the in practice more useful expression
\begin{equation}
\label{MaxwellRep2}
f(\theta,\phi) = C(\mathbf{e}(\theta,\phi)\cdot \mathbf{v}_1)\ldots(\mathbf{e}(\theta,\phi)\cdot \mathbf{v}_l) + r^2F(\theta,\phi),
\end{equation}
where $\theta$ and $\phi$ 
describe the sphere in spherical coordinates and 
$\mathbf{e}(\theta,\phi) 
= (x(\theta,\phi),y(\theta,\phi),z(\theta,\phi))/r(x,y,z)$, and $F$ is a homogeneous polynomial in the variables $x,y,z$ of degree $\leq l-2$.
Due to the fact that spherical harmonics provide a basis for harmonic functions, each multipole of a spherical harmonic decomposition of CMB fluctuations on the sky can be identified uniquely with a set of MPVs.

\subsection{Extraction of multipole vectors}
\label{approaches}

There exist several approaches to MPVs and their calculation from a spherical harmonic decomposition, three of which we will review briefly in the following. 
While the approach via coherent states appears to be best suited for the investigation of statistical properties, in this work the tensorial approach has been 
used to calculate the MPVs numerically \cite{CopiProg}.

\subsubsection{Tensorial construction}
\label{tensorial}

Copi \textit{et al.} first applied the long-forgotten method of MPVs to the analysis of CMB data in \citep{MultiCopi}.

Let the fragments $f_l$ be as in Eq.~\eqref{fl_def}. They are harmonic and homogeneous polynomials of degree $l$ in $x,y,z$ and thus $f_l(x,y,z) = F_{i_1\cdots i_l}e^{i_1}\ldots e^{i_l}$. 
In order to guarantee the uniqueness of this expression, it is inevitable to impose further restrictions on the coefficients $F_{i_1\cdots i_l}$, which can be regarded as coefficients of a tensor $F$, and on the product $e^{i_1}\ldots e^{i_l}$. Both factors have to be trace-free and symmetric:
\begin{equation}
\label{fl}
f_l(\mathbf{e}) = F^{(l)}_{i_1\cdots i_l}[ e^{i_1}\cdots e^{i_l} ]  =: A^{(l)} \left[ v_{i_1}^{(l,1)}\cdots v_{i_l}^{(l,l)} \right] [e^{i_1}\cdots e^{i_l}].
\end{equation}
The brackets denote the symmetric trace-free part of the interior.
Equation \eqref{fl} defines the MPVs, which can be calculated uniquely, up to rescaling, from the spherical harmonic data. One recovers $F$ from $f_l$ via integration
\begin{equation}
F_{i_1\cdots i_l}^{(l)} = \frac{(2l+1)(2l)!}{(4\pi)2^l(l!)^2}\int_{\mathcal{S}^2} \! \mathrm{d}\mathbf{e} \, f_l(\mathbf{e})\left[ e^{i_1} \cdots e^{i_l} \right],
\end{equation}
and afterwards peels off the first MPV by writing
\begin{equation}
\label{peeloff}
F^{(l)}_{i_1\cdots i_l} = \left[ v^{(l,1)}_{i_1} a^{(l,1)}_{i_2\cdots i_l} \right],
\end{equation}
where $a^{(l,1)}$ is a rank $l-1$ tensor. In the same manner one can peel off the second MPV from $a^{(l,1)}$ leaving a rank $l-2$ tensor. Repeating this procedure until a rank 1 tensor is left yields all $l$ MPVs. By performing a more detailed mathematical calculation one can write down a system of equations that relates the $a_{lm}$ and the $\mathbf{v}^{(l,j)}$; for more details see \cite{MultiCopi}. Copi's MPV calculation program \cite{CopiProg}, which was used by the authors, evaluates this system and returns the MPVs.

Finally note that no information is lost in the transition from spherical harmonics to MPVs. For each $l$ there are $2l+1$ real degrees of freedom in the spherical harmonic decomposition, namely the real and imaginary parts of all $a_ {lm}$ with $m\geq 0$. On the other hand, $l$ unit vectors and an amplitude constitute as well $2l+1$ real degrees of freedom since due to the normalization condition a single unit vector in $\mathbb{R}^3$ has $2$ degrees of freedom, and the amplitude is just a scalar which contributes one further degree of freedom.

\subsubsection{Algebraic construction}
\label{algebraic}

Katz and Weeks \cite{KatzWeeks} applied B\'{e}zout's theorem from algebraic geometry 
to proof Sylvester's theorem. The advantage of this approach is its mathematically sophisticated nature. Furthermore, like the tensorial approach, it yields an iterative prescription calculating the MPVs, and even an explicit expression for the residual polynomial $F$ can be obtained. This approach has a long history    dating back to Hilbert and Courant, see \cite{Hilbert}.

A homogeneous polynomial $P$ of degree $l$ on $\mathbb{R}^3$ may be written uniquely up to reordering and rescaling as
\begin{align}
\label{Theorem1}
\begin{split}
P(x,y,z) = & \lambda (a_1 x+b_1 y+c_1 z)\cdots \\
 &\cdots (a_l x+ b_l y+ c_l z) \\
& + (x^2+y^2+z^2)R,
\end{split}
\end{align}
where $R$ denotes a residual polynomial which is homogeneous of degree $l-2$. If $l<2$ one sets $R \equiv 0$, and for $l=2$ the zero can be replaced by a nonvanishing constant.

Let now $f$ be an arbitrary, especially not necessarily homogeneous, polynomial of degree $l$ restricted to the two-sphere. It can be written as a sum of homogeneous polynomials of degree $i$, $f_i$, via $f = \sum_{i=0}^l f_i$. According to \eqref{Theorem1}, up to reordering and rescaling $f_i$ can be decomposed into linear factors and a residual polynomial $f_i = \lambda_i \prod_{j=1}^i(v^{(l,j)}_x x+v^{(l,j)}_y y+v^{(l,j)}_z z) + R_{i-2}(x,y,z)$. Since $R_{i-2}$ is homogeneous of degree $i-2$, the sum $f'_{i-2}:=f_{i-2} + R_{i-2}$ is again homogeneous of degree $i-2$. Applying \eqref{Theorem1} recursively on the rest of the sum eventually results in 
\begin{equation}
\label{Corollary2}
f(x,y,z) = \sum_{i=0}^l \lambda_i \prod_{j=1}^i (v^{(i,j)}_x x+v^{(i,j)}_y y+v^{(i,j)}_z z) \quad \text{on } \mathcal{S}^2.
\end{equation}
The scalar product of MPVs with the unit vector in the $(\theta,\phi)$-direction is given by the $i=l$-term 
while the rest of the sum constitutes the residual polynomial $F$. For more details we refer to \cite{KatzWeeks}.

\subsubsection{Construction via Bloch coherent states}
\label{majorana}

Dennis used a very physical approach to proof Sylvester's theorem and associate MPVs to 
spherical harmonics \citep{Dennis2004}. A complex spin-state in nonrelativistic one-particle 
quantum mechanics with spin $1/2$ can be represented via one point on the two-sphere. 
This concept is known as the Bloch sphere. Extending this concept to higher integer 
spins and assuming the state is real yields Sylvester's theorem. The big advantage of this 
approach is the capability of calculating a joint probability density for the MPVs using techniques 
from random polynomial theory. An analytic result for the joint probability density in principal 
allows us to compute confidence levels for certain CMB statistics analytically. 

A similar approach, but with slightly different focus, was used in \cite{Helling}. By rotating the 
highest weight spin state one receives Bloch coherent states, i.e.\ those coherent 
states associated to $\mathrm{SO}(3)$, and the overlap of such a coherent state at rotation 
angles $\theta$ and $\phi$ with a normalized spin state gives, after stereographic projection,
the Majorana polynomial below. Using the Bloch states one can define an extended version 
of the von Neumann entropy, called Wehrl entropy, which measures quantum randomness.

The formalism below has already been used by Schupp in 1999 in the proof of some special 
cases of Lieb's conjecture for the Wehrl entropy of Bloch coherent states; see \cite{Schupp}.

Let $|\Psi \rangle$ denote a quantum mechanical state with definite integer spin $l$, i.e., an eigenstate of the total angular momentum operator $\hat{L}^2$. It corresponds to a harmonic function in the language of the previous subsections. The eigenstate property allows us to expand the state in terms of eigenstates of the $z$-component of the angular momentum operator $\hat{L}_z$
\begin{equation}
|\Psi\rangle = \sum_{m=-l}^l \Psi_m |m,l\rangle, \quad \Psi_m \in \mathbb{C},
\end{equation}
which in position space is nothing other than the spherical harmonic decomposition.
Let $\hat{R}_z(\phi)$ denote the operator which executes a rotation by the angle $\phi$ around the $z$-axis and $\hat{R}_y(\theta)$ the rotation by $\theta$ around the $y$-axis and define 
\begin{align}
|m,l;\theta,\phi\rangle := \hat{R}_z(\phi)\hat{R}_y(\theta)|m,l\rangle.
\end{align}
This is an eigenstate of the $\mathbf{e}(\theta,\phi)$-parallel component of $\hat{L}$. The spherical harmonics are then given by
\begin{equation}
Y_{lm}(\theta,\phi) = \sqrt{\frac{2l+1}{4\pi}}\langle 0,l;\theta,\phi|m,l\rangle.
\end{equation}
The state $|\Psi\rangle$ can now be expressed via spherical harmonics by projecting on the rotated $m=0$ state
\begin{equation}
\Psi(\theta,\phi):= \langle 0, l;\theta,\phi|\Psi\rangle = \sqrt{\frac{4\pi}{2l+1}}\sum_{m=-l}^l \Psi_m Y_{lm}(\theta,\phi).
\end{equation}
After stereographic projection from the south pole
\begin{equation}
\label{stereographic}
\zeta(\theta,\phi) = \tan(\theta/2)\exp(i\phi),
\end{equation}
and using some group theory, the spin spate $|\Psi\rangle$ can be decomposed according to the $SL(2,\mathbb{C})$ basis functions $\mu_{k-l}\zeta^k$ with $k\in\mathbb{N}_0$
\begin{equation}
\label{Majorana function}
f_{\Psi}(\zeta) := \langle -l,l;\zeta|\Psi\rangle = \frac{\exp(-\mathrm{i}l\arg(\zeta))}{(1+|\zeta|^2)^l}\sum_{m=-l}^l \Psi_m \mu_m \zeta^{l+m},
\end{equation}
with the numerical factor $\mu_m = (-1)^{l+m}\sqrt{\binom{2l}{l+m}}$. The representation \eqref{Majorana function} of the state is called the Majorana function. It is a product of a $\zeta$-dependent factor and a polynomial of degree $2l$ in $\zeta$ which contains all the information about the original state. This polynomial is called the Majorana polynomial and it determines the roots of the Majorana function. Since it is a polynomial of degree $2l$ in the complex variable $\zeta$, it possesses $2l$ complex roots according to the fundamental theorem of algebra, and therefore it can be factorized
\begin{equation}
f_{\Psi}(\zeta) = \frac{\exp(-\mathrm{i}l \arg(\zeta))}{(1+|\zeta|^2)^l}(-1)^{2l}\Psi_{m=l} \prod_{n=1}^{2l}(\zeta-\zeta_n).
\end{equation}
The $2l$ roots can be backprojected onto the Riemannian sphere $\mathcal{S}^2 \, \tilde{=} \, \hat{\mathbb{C}} \, \tilde{=} \, \mathbb{C} \cup \{ \infty \}$. These backprojected roots $\mathbf{v}(\zeta_n)$ are called Majorana vectors. In the case of a real $\Psi(\theta,\phi)$ they are identical to the MPVs
\begin{equation}
\mathbf{v}^{(l,j)} \equiv \mathbf{v}(\zeta_j)
\end{equation}
for a given $l$.

The Majorana function of the rotated state obeys $f_{\hat{R}_{v,\theta}\Psi}(\zeta) = f_{\Psi}(MT(\zeta))$, where $\mathrm{MT}$ denotes a unitary M\"{o}bius transformation. Consequently its zeros also transform under a unitary M\"{o}bius transformation. After backprojection this transformation corresponds to a rotation through $SO(3)$. Majorana vectors rotate rigidly like ordinary three-vectors. 

A further property of Majorana vectors is their appearance in antipodal pairs if the original state is real
\begin{equation}
f_{\Psi}(-1/\zeta^{*}) = f_{\Psi}(\zeta)^{*}.
\end{equation}
Hence, $\zeta$ is a root of the Majorana function if and only if $-1/\zeta^{*}$ is a root, but $-1/\zeta^{*}$ is the image under the stereographic projection of the antipode of the Majorana vector determined by $\zeta$. This property does not hold if the original state is complex. Complex functions on the sphere cannot be represented by $l$ MPVs.

There have been several further approaches to MPVs, for example by investigating their topological implications in \cite{Arnold}.

\section{Statistical properties of multipole vectors}
\label{Stat_props}

The spherical harmonic coefficients $a_{lm}$ of the CMB temperature fluctuations are attached with 
a notion of randomness implied by inflationary fluctuations. Standard inflationary scenarios lead to 
Gaussianity of these coefficients. Whether they are really Gaussian or not, they definitely constitute 
a set of random variables. The MPVs, which depend only on these coefficients, inherit the 
randomness from these coefficients. One may now ask what kind of probability distribution 
they obey exactly. 

Dennis and Land attended to this question first in \cite{Dennis2005}, followed up by \cite{Dennis2007}. 
For this purpose the coherent state approach turns out to be especially useful because MPVs are the 
roots of a complex polynomial whose coefficients are the $a_{lm}$ times some numerical factor. 
Therefore we have to deal with the probability density of roots of random polynomials which is currently a much studied field of statistical mathematics.

In Appendix \ref{RMT} we present some first ideas on how to apply results from random matrix theory and the theory of 
Gaussian analytic functions to the problem of the joint probability distribution of MPVs. Future advances in this direction could allow for determining p-values with arbitrary precision in short computing time.

This section is intended to provide a review of the derivation of the MPV joint probability distribution. 
The essential properties are the statistical decoupling of MPVs at different angular scales and the 
nontrivial correlation between MPVs at a given angular scale $\pi/l$. Furthermore, it is important to note 
that even if the underlying temperature fluctuation field is Gaussian, the MPV distribution is not and 
hence it is not enough to consider only one- and two-point functions, but one needs the full set of all 
$n$-point functions, where $n=1,\ldots,l$ for a given $l$. Although an explicit expression for the probability 
distribution of the MPVs has been found before in \cite{Dennis2005}, it turns out to be of limited use for 
practical purposes, except for the lowest multipoles $l=1,2,3$. In this work we use Monte Carlo methods, 
which appear to yield results faster than numerical integration of the analytic expression. Nevertheless, the following review yields a solid understanding of what kind of behavior one should expect.

\subsection{Isotropy and Gaussianity}

Let us first focus on the description of isotropy and Gaussianity in the CMB data and the 
difference between both. 

The temperature fluctuation field is Gaussian if for all $n \in \mathbb{N}$ and all $\mathbf{e}_i \in \mathcal{S}^2$ with $i=1,\ldots,n$ the probability distribution of $\delta T/T_0$ follows
\begin{equation}
p(\delta T/T_0) = \frac{1}{\mathcal{N}}\exp\left( -\frac{1}{2}\sum_{ij}\left( \frac{\delta T}{T_0} \right)_i (D^{-1})_{ij} \left( \frac{\delta T}{T_0} \right)_j \right),
\end{equation}
with $(\delta T/T_0)_i = \delta T(\mathbf{e}_i)/ T_0 $ and some normalization constant $\mathcal{N}$. The matrix $D$ is the correlation matrix $D_{ij} = \langle (\delta T/T_0)_i (\delta T/T_0)_j \rangle$. The Gaussianity of $\delta T$ implies Gaussianity of the $a_{lm}$ that obey
\begin{equation}
p(\{ a_{lm} \}) = \frac{1}{\mathcal{N}'} \exp\left( -\frac{1}{2}\sum_{l,l',m,m'} a_{lm}^* (C^{-1})_{lml'm'}a_{l'm'} \right),
\end{equation}
with $C_{lml'm'} = \langle a_{lm}^* a_{l'm'} \rangle$ and $\mathcal{N}'$ some normalization constant which is in general different from $\mathcal{N}$. A Gaussian field is fully characterized by its correlation matrix and if we demand isotropy additionally, then $C_{lml'm'} = \delta_{mm'}\delta_{ll'} C_l$. In this case we have
\begin{equation}
\label{GaussianAlm}
p(\{a_{lm}\}) = \prod_{lm}\frac{\exp(-|a_{lm}|^2/(2C_l))}{\sqrt{2\pi C_l}},
\end{equation}
i.e., the $a_{lm}$ are identically and independently distributed complex Gaussian random variables with variance $C_l$, or alternatively all real and imaginary parts $\Re a_{lm}$, $\Im a_{lm}$ as well as $a_{l0}$ are identically and independently distributed real Gaussian random variables with variance $C_l$.

Isotropy and Gaussianity do not necessarily imply each other. Consider for example a distribution which is gained by an isotropic and Gaussian distribution via introducing a cutoff for large values of $|a_{lm}|$. By this we mean $p(a_{lm}) = 0$ if $|a_{lm}| > \kappa \in \mathbb{R}$ for all $l$ and $m$. This distribution is not fully Gaussian any longer but does not lose its isotropy. On the other hand a general Gaussian distribution does not need to be isotropic.

\subsection{Probability distribution}

\begin{figure}
\resizebox{\hsize}{!}{\includegraphics{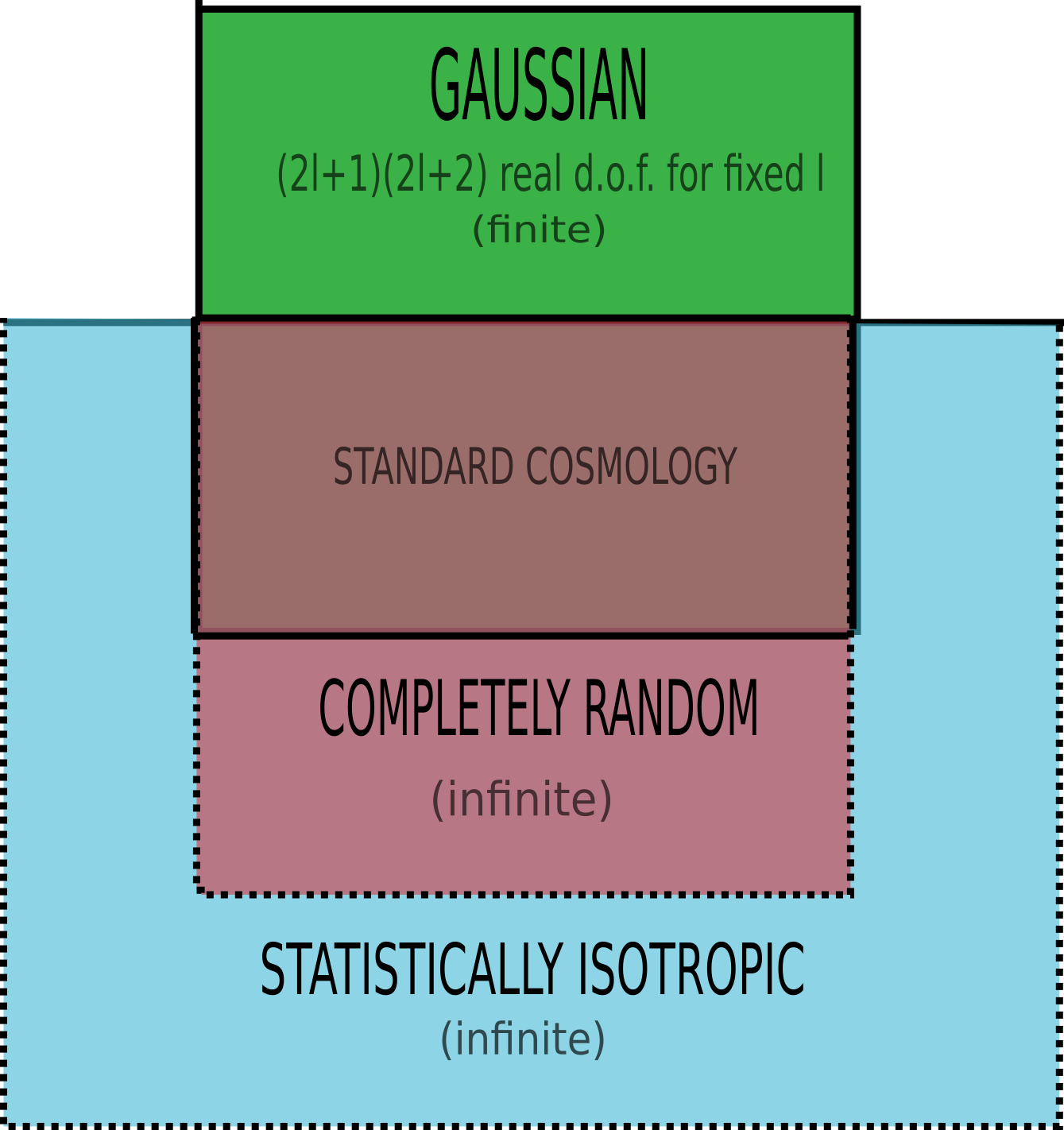}}
\caption[Visualization of relationship among distributions]{Visualization of the relation between Gaussianity, statistical isotropy and complete 
randomness. Dotted lines denote infinite extension.}
\label{distributions}
\end{figure}

It turns out that the joint probability density for the MPVs of fixed angular momentum $l$ is the same for all so-called completely random sets of $a_{lm}$. This means that the probability density of the coefficients, $p(a_{lm})$, depends only on the sum $\sum_m |a_{lm}|^2$, respectively on the power spectrum estimator $\hat{C}_l$, see \cite{Dennis2005} and \cite{Dennis2007}. An isotropic and Gaussian distribution is of course included in the set of completely random distributions, see Fig.~\ref{distributions}. Note that complete randomness does not require the $a_{lm}$ to be statistically independent. Rather, if they are statistically independent and completely random, they automatically have to be Gaussian. Gaussianity itself in combination with complete randomness implies statistical isotropy, but not vice versa, and therefore we shall insist on complete randomness for the rest of this publication and treat it as a basic assumption which incorporates statistical isotropy if Gaussianity is given. The most important class of distributions for cosmology is given by the intersection of completely random and Gaussian $a_{lm}$, which we call ``standard cosmology'' in Fig.~\ref{distributions}. Since no sizable deviations from Gaussianity have been observed so far, see e.g. \cite{KomatsuNonGaussianity} or \cite{Planck2015XVII}, assuming that Gaussianity holds true allows for investigating isotropy solely.

For fixed $l$ the set of Gaussian distributions is described by a finite number of degrees of freedom, since due to Wick's theorem the expectation values $\langle a_{lm} \rangle$ and the two-point functions $\langle a_{lm} a_{lm'}^* \rangle$ uniquely determine the distribution. Respecting the reality condition $a_{lm}^* = (-1)^m a_{l,-m}$ yields $2l+1 + (2l+1)^2 = (2l+1)(2l+2)$ real degrees of freedom. The sets of statistically isotropic as well as completely random distributions are \textit{a priori} not bounded in their degrees of freedom. Any $n$-point function, with $n \in \mathbb{N}$, contributes to the knowledge of the distribution. Hence both distributions have at least a countably infinite set of degrees of freedom. The intersection of the completely random case and the Gaussian case coincides with the intersection of the isotropic and Gaussian case. Nevertheless, there exist completely random  $a_{lm}$ which are not Gaussian, as for example a delta distribution $\delta(\hat{C}_l)$. In principle there can also exist statistically isotropic, non-Gaussian distributions which are not completely random. For this one just forms some rotationally invariant quantity $\mathcal{Q}$ --which shall not be a function of the $C_l$-- out of the $a_{lm}$ and considers its distribution $p(\mathcal{Q})$. That completely random distributions form a subset of statistically isotropic ones can be seen as follows: since a rotation $\hat{R}$ acts on $\mathbb{C}^{2l+1}$ as a special unitary transformation, its determinant vanishes and therefore 
\begin{equation}
\langle \prod_i \mathcal{O}_i(\hat{R}(\mathbf{e}_i)) \rangle = \int \prod_m \mathrm{d}a_{lm} \prod_i \mathcal{O}_i(\{ a_{lm} \})p(\{ \hat{R}^{-1} a_{lm} \}).
\end{equation}
The property of statistical isotropy reduces to the rotational invariance of the joint probability.
In the completely random case we have $p(\{a_{lm}\}) = p(\sum_m |a_{lm}|^2)$. Let the unitary operator corresponding to $\hat{R}^{-1}$ acting on $\mathbb{C}^{2l+1}$ be denoted by $\mathcal{D}$ (which is related to Wigner's D-matrix; see \cite{Wigner}) and write the set of $a_{lm}$ for fixed $l$ as a vector $\mathbf{a}_l \in \mathbb{C}^{2l+1}$; then 
\begin{align}
\hat{R}\left( \sum_m |a_{lm}|^2 \right) &= \sum_m |\hat{R}(a_{lm})|^2 = (\mathcal{D}\mathbf{a}_l)^{\dagger}\cdot (\mathcal{D}\mathbf{a}_l) \nonumber \\
&= \sum_m |a_{lm}|^2,
\end{align}
due to the unitary representation of $SO(3)$ as $SU(2)$. Hence, completely random sets of $a_{lm}$ always obey statistical isotropy.

For CMB analysis one needs the joint probability distribution of MPVs because inside one multipole they are not independent of each other. This stems directly from the behavior of random roots which tend to repel each other. Contrary to this, the MPVs from different multipoles are perfectly independent.

The first calculation of the joint probability densities of random spin-$l$ states was performed in 1995 by Hannay, see \cite{Hannay}. He notes that the Majorana function equals exactly the Bargmann function of the spin state in the Segal-Bargmann space \cite{Bargmann}. This representation of quantum states can be seen as a third leg of standard quantum mechanics accompanying Heisenberg's matrix- and Schr\"{o}dinger's wave function quantum mechanics.

It turns out that in the completely random case the $n$-point density can be written as a normalized permanent
\begin{equation}
\label{PDF general}
\begin{split}
p_n^l&(\zeta_1,\ldots,\zeta_n, \zeta_{l+1}=-1/\zeta_1^*,\ldots,\zeta_{l+n}=-1/\zeta_n^*) \\
& = \frac{1}{\pi^n} \frac{\mathrm{per}(C-B^{\dagger}A^{-1}B)}{\det(A)},
\end{split}
\end{equation}
with
\begin{align}
f_i &:= f_{\Psi}(\zeta_i) \quad \text{\textit{Majorana function evaluated at the root}} \nonumber\\
\label{A}
A_{ij} &= \langle f_if_j^* \rangle \nonumber \\
&\stackrel{\mathrm{isotropy}}{=} \sum_{m,m' = -l}^l (-1)^{m+m'} \left[\binom{2l}{l+m} \binom{2l}{l+m'} \right]^{1/2} \nonumber \\  
& \cdot \underbrace{\langle a_{lm}a^*_{lm'}\rangle}_{=C_l\delta_{mm'}}\zeta^{l+m}_i (\zeta^*_j)^{l+m'} \nonumber \\
 &= C_l(1+\zeta_i\zeta_j^*)^{2l} \\
 \label{B}
 B_{ij} &= \langle f_i f_j'^* \rangle \stackrel{\mathrm{isotropy}}{=} C_l 2l \zeta_i(1+\zeta_i\zeta_j^*)^{2l-1} \\
 \label{C}
 C_{ij} &= \langle f_i' f_j'^* \rangle \stackrel{\mathrm{isotropy}}{=} C_l 2l (1+2l \zeta_i \zeta_j^*)(1+\zeta_i\zeta_j^*)^{2l-2} 
,
\end{align}
where the second equalities only hold in the isotropic case; see also \cite{Dennis2007}. The matrices $A,B,C$ are $(2n\times 2n)$-matrices.
Calculating $A^{-1}$ and inserting the explicit formulas from \eqref{A} to \eqref{C} yields the joint probability density of $1 \leq n \leq l\,$ MPVs. Note that here the function $f$ can in principle be complex. Isotropy and Gaussianity enter the game when the precise expressions \eqref{A}-\eqref{C} are inserted. Even though the function $f$ can be complex, one should remember that the representation of the function by MPVs is possible only for real functions.

An alternative derivation of the full joint density ($n=l$), made by Dennis in \cite{Dennis2007}, uses the fact that the coefficients of any polynomial can be expressed by a symmetric polynomial of its roots leading to the full joint density
\begin{align}
\label{PDF 2l}
\begin{split}
p_{l}^l(\{ \zeta_i \}) = & \frac{(2l-1)!!\prod_{j=1}^{2l} j!}{(2\pi)^l l! \prod_{j=1}^l |\zeta_j|^2} \\
 & \cdot \frac{\prod_{j,k=1,j<k}^{2l} |\zeta_j-\zeta_k|}{\left( \sum_{\sigma \in S_{2l}} \prod_{j=1}^{2l} (1+\zeta_j \zeta_{\sigma(j)}^*) \right)^{l+1/2}}.
 \end{split}
\end{align}
When projecting back to the sphere, the Jacobi determinant for this transformation has to be further taken into account. Note that $f$ has to be real for Eq.~\eqref{PDF 2l} to be valid, since the result was obtained by implicitly setting $\zeta_{l+i} = -1/\zeta_i^*$ which is only true for real Majorana polynomials.

In the case $n=1$ the distribution of MPVs on one hemisphere of the two-sphere simplifies to a uniform distribution
\begin{equation}
\label{p1}
p_1^l([\mathbf{v}]) = p(\theta,\phi) =  \frac{1}{2\pi}
\end{equation}
according to the surface measure. So when drawing MPVs from an ensemble of $a_{lm}$, the first MPV one draws is always uniformly distributed. For a derivation of \eqref{p1} see Appendix \ref{one-point}.

\begin{figure}
\resizebox{\hsize}{!}{\includegraphics{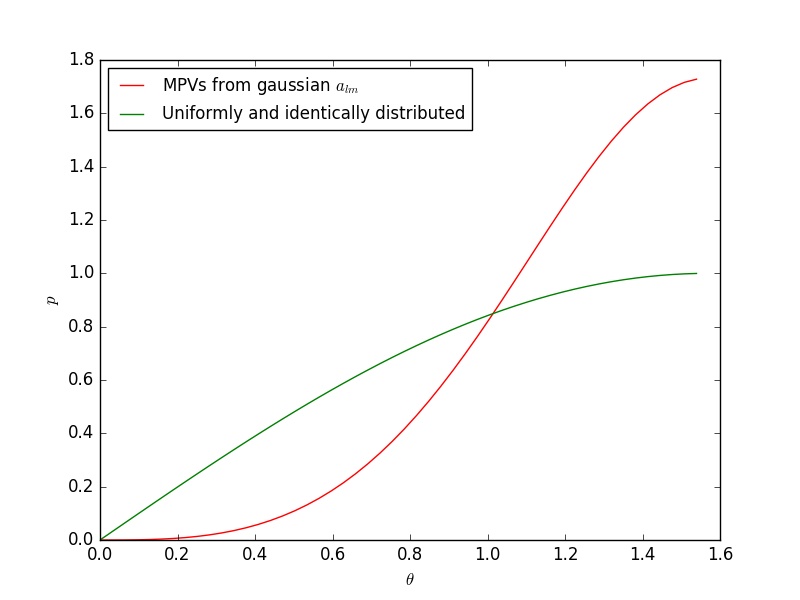}}
\caption[2-point density for $l=2$: MPV vs. uniform]{Comparison of the two-point density of MPVs drawn from a Gaussian ensemble of $a_{lm}$ and identically uniformly distributed pseudo-MPVs for $l=2$}
\label{MPV vs Uni}
\end{figure}

Since MPVs rotate rigidly their density depends only on the relative angle. Thus, for $l=n=2$ one of the two vectors can be fixed to an arbitrary direction, without loss of generality to the north pole, and the second one encloses an angle $\Theta$ with the first one. The two-point density is then given by \citep{Dennis2007}
\begin{equation}
\label{2-point}
p_2^2(\Theta) = \frac{27 \sin^3(\Theta)}{(3+\cos^2(\Theta))^{5/2}}.
\end{equation}
For comparison imagine a world in which the MPVs are uniformly and independently distributed on the upper hemisphere. Then the first MPV can again be fixed to the north pole and the second is still uniformly distributed on the upper hemisphere. In this case the two-point density would be $p_{\mathrm{uni}}(\Theta) = \sin(\Theta)/2$. Both probability densities are normalized in a way such that $\int_0^{\pi/2}\! \mathrm{d}\Theta \, p(\Theta) = 1$.
In Fig.~\ref{MPV vs Uni} the two densities are plotted together. One can see that the interaction of MPVs leads to repulsion. Bigger angles are more probable in the case of real MPVs than in the case of uniformity and independence. Such a behavior is characteristic for roots of random polynomials.

The results from \eqref{PDF general} are \textit{a priori} complicated expressions, even for the case $n=3$, whose integration does not allow for a faster numerical computation of confidence levels than a full Monte Carlo simulation.

\section{CMB data}
\label{data}

For our analysis we make use of the Planck 2015 full-sky CMB intensity maps (see
\footnote{\url{http://irsa.ipac.caltech.edu/data/Planck/release\_{}2/all-sky-maps/matrix\_{}cmb.html}}) with 
$N_{\mathrm{side}} = 2048$ and treat it within HEALPy, 
which is a HEALPix \footnote{\url{http://healpix.sf.net}} implementation for \textsc{python}, see
\cite{HealpixPaper}.

The Planck full-sky maps are provided in nine different frequency bands. These maps have been foreground cleaned to produce best estimates of intensity and 
polarization of the measured CMB signal from the sky after instrumental and known systematical effects, like the dipole and quadrupole (DQ) induced by the motion 
of the Sun and Earth with respect to the cosmic frame; the light from the zodiac cloud; and most important galactic foregrounds (synchrotron radiation, free-free emission, 
thermal dust, CO lines, anomalous microwave radiation) have been removed. Using four different cleaning algorithms, foreground cleaned full-sky maps of CMB 
temperature intensity are constructed; for the details of the component separation process we refer to \cite{Planck2015IX}. These different cleaning algorithms, are 
called COMMANDER, NILC, SEVEM and SMICA.

The four maps have been used to extract the MPVs up to $l=50$ by using a tensorial algorithm \cite{CopiProg}, see also Sec.~\ref{tensorial}. A list of the MPVs for 
the multipoles $l=2,3,4,5$ can be found in Tab.~\ref{MPV_tab}; text files containing the MPVs for higher $l$ are provided at \url{https://github.com/MPinkwart/MPV-files-Pinkwart-Schwarz}.

\begin{table}
\begin{tabular}{|c|c|c|c|c|}
\hline
$l$ & COMMANDER & NILC & SEVEM & SMICA  \\
\hline \hline
$2$ & (6.9, 21.2) & (12.8, 20.3) & (13.3, 26.2) & (5.7, 23.7)  \\
 & (119.1, 18.2) & (117.5, 20.1) & (83.6, 12.3) & (121.5, 21.9) \\
\hline
$3$ & (25.6, 8.8) & (23.2, 9.3) & (33.0, 5.4) & (22.1, 8.8)  \\
 & (86.3, 38.4) & (86.8, 37.7) & (61.3, 35.0) & (88.1, 38.8)  \\
 & (317.7, 5.0) & (315.3, 7.9) & (140.5, 0.6) & (314.8, 10.5)  \\
\hline
$4$ & (69.5, 3.2) & (69.9, 4.5) & (71.1, 19.2) & (68.8, 2.2) \\
 & (207.5, 72.6) & (203.9, 70.5) & (189,4, 73.0) & (207.8, 38.5)  \\
 & (211.2, 36.7) & (212.6, 40.0) & (201.7, 38.0) & (214.8, 69.9)  \\
 & (333.4, 29.1) & (333.5, 27.7) & (336.5, 27.3) & (335.9, 26.1)  \\
\hline
$5$ & (43.3, 33.3) & (44.2, 35.8) & (51.5, 28.2) & (44.3, 36.7) \\
 & (98.7, 35.7) & (96.8, 36.1) & (79.8, 35.3) & (98.3, 36.2)  \\
 & (174.0, 3.9) & (175.2, 3.4) & (174.2, 3.8) & (175.7, 3.3)  \\
 & (232.1, 54.2) & (232.4, 55.3) & (232.1, 58.2) & (234.9, 55.3)  \\
 & (287.4, 31.6) & (286.1, 31.6) & (290.1, 18.9) & (285.2, 32.8)  \\
\hline
\end{tabular}
\caption{MPVs from $l=2$ to $l=5$ in galactic coordinates $(l,b)$ in deg with the precision of one decimal. All MPVs have been taken to lie in the northern hemisphere and for a given multipole they were ordered according to their value of the galactic longitude. MPVs in the same line cannot necessarily be identified with each other since they are not \textit{a priori} ordered inside of a given multipole. 
}
\label{MPV_tab}
\end{table}

\section{Statistics and Simulations}
\label{Analysis}

\subsection{Statistics}

One needs statistics that deliver information about both intramultipole alignments and alignments of multipoles with some given astrophysical direction, which in the following will be referred to as the outer direction. Furthermore, possible statistics are not allowed to depend on the ordering of MPVs for a fixed $l$ since this ordering is completely arbitrary and contains no information. Additionally, the statistics may not depend on the hemisphere, since MPVs are lines rather than vectors. Eventually, they may not be sensitive to the equator since the gluing mechanism at the equator should be hidden. In the following, intramultipole statistics are sometimes also referred to as inner statistics and statistics that investigate correlations with outer directions as outer statistics. 

We used the following two outer statistics
\begin{align}
S^{||}_{\mathbf{D}}(l):= & \frac{1}{l}\sum_{i=1}^l |\mathbf{v}^{(l,i)} \cdot \mathbf{D}| \\
S^{v}_{\mathbf{D}}(l):= & \frac{2}{l(l-1)} \sum_{1 \leq i < j \leq l} |(\mathbf{v}^{(l,i)} \times \mathbf{v}^{(l,j)})\cdot \mathbf{D}|, 
\end{align}
where $\mathbf{v}^{(l,i)}$ denotes the $i$th MPV belonging to multipole $l$ and $\mathbf{D}$ some outer direction, which will be specified in Sec.~\ref{data}.

Furthermore, we use two inner statistics
\begin{align}
S^{||}(l) := & \frac{2}{l(l-1)} \sum_{1 \leq i < j \leq l} |\mathbf{v}^{(l,i)} \cdot \mathbf{v}^{(l,j)}|  \\
S^{v}(l) := & \frac{6}{l(l-1)(l-2)} \sum_{1 \leq i < j < k \leq l} | (\mathbf{v}^{(l,i)} \times \mathbf{v}^{(l,j)}) \cdot \mathbf{v}^{(l,k)} | .
\end{align}

All statistics are normalized such that they take values in the unit interval $[0,1]$. Each summand in every statistic ranges from $0$ to $1$ while the number of summands is $l = \binom{l}{1}$ for $S^{||}_{\mathbf{D}}$, $l(l-1)/2 = \binom{l}{2}$ for $S^{||}$ and $S^{v}_{\mathbf{D}}$, and $l(l-1)(l-2)/6 = \binom{l}{3}$ for $S^{v}$. 

The statistic $S^{||}_{\mathbf{D}}$ measures the alignment of a multipole with an outer direction while $S^{v}_{\mathbf{D}}$ measures the orthogonality of a multipole with respect to this outer direction. The statistic $S^{||}$ measures the possible linearity of the respective multipole itself while $S^{v}$ measures possible planarity.

Let $X^{l,i} := |\mathbf{v}^{(l,i)} \cdot \mathbf{D}|$, then the expectation value of $S^{||}_{\mathbf{D}}$ is
\begin{equation}
\langle S^{||}_{\mathbf{D}}(l) \rangle =  \frac{1}{l}\sum_{i=1}^l \langle X^{l,i} \rangle  = \frac{1}{2}.
\end{equation}
This result holds for all types of completely random $a_{lm}$. Due to the correlation of MPVs inside one multipole the variance of $S^{||}_{\mathbf{D}}$ is
\begin{equation}
\mathrm{Var}(S^{||}_{\mathbf{D}}(l)) = \frac{1}{l^2}\left( \frac{l}{12} - \frac{l(l-1)}{4} + 2 \sum_{1 \leq i  j \leq l} \langle X ^{l,i} X ^{l,j} \rangle \right),
\end{equation}
where for the calculation of $\langle X^{l,i}X^{l,j} \rangle$ one uses the two-point density \eqref{2-point}, yielding
\begin{equation}
\label{theodev}
\mathrm{Var}(S^{||}_{\mathbf{D}}(2)) = \frac{1}{4}\left( -\frac{1}{3} + \frac{2}{3}\left( 6 - \frac{10}{\sqrt{3}} \right)\pi \right) \approx 0.035 .
\end{equation}
If the MPVs are not correlated but are all independent, $S^{||}_{\mathbf{D}}$ would follow a slightly modified Irwin-Hall distribution, see \cite{Irwin, Hall},
\begin{equation}
p_{S^{||}_{\mathbf{D}}(l)}(s) = \frac{l}{2(l-1)!}\sum_{k=0}^l (-1)^k \binom{l}{k}(ls-k)^{l-1} \mathrm{sgn}(ls-k),
\end{equation}
which would result in a variance of $S^{||}_{\mathbf{D}}(2)$ of about $0.042$ which is slightly larger than the variance in the completely random case, showing again that the intramultipole correlation tightens confidence regions. 
Note that in our analysis we do not use analytical results, since the numerical computation of confidence levels using the full joint probability \eqref{PDF 2l} turns out to be numerically more demanding than a simple Monte Carlo simulation. 

In order to compare the analytical results with the numerics, we consider $1000$ maps from isotropic and Gaussian random $a_{lm}$ and compare their MPV statistics with the one from the cleaned Planck maps. From Fig.~\ref{S1_dip_small} one deduces that the theoretical result \eqref{theodev} for the variance is compatible with the numerical result for the $1\sigma$-region because $\sqrt{0.035} \approx 0.187$.

To characterize and quantify a possible violation of the completely random hypothesis, we introduce a notion of likelihood suggested in \cite{Abramo}. Let us first define what we mean by the p-value: let $S_{i,l}$ be the data point of statistic $S_i(l)$ received by one of the four Planck maps, where $i \in \{1,2,3,4 \}$ runs through the four statistics. Define the p-value of this data point as
\begin{equation}
P(S_{i,l}) := \int_0^{S_{i,l}} \! \mathrm{d}s \, p_{S_i(l)}(s)
,
\end{equation}
i.e. small ($\ll 1$) as well as large ($\approx 1$) p-values indicate unusual behavior. Now let $S_{i,l}^M$ denote a data point as above received from map $M$ (COMMANDER, NILC, SEVEM or SMICA). We define the outer likelihood
\begin{equation}
\label{Louter}
L_{l,\mathbf{D}}^{\mathrm{outer}}(M) := 4^2 \prod_{\text{outer}} P\left( S_{l,i}^M\right)\left(1-P\left(  S_{l,i}^M\right)\right) ,
\end{equation}
as well as the inner likelihood
\begin{equation}
\label{Linner}
L_l^{\mathrm{inner}}(M) := 4^2 \prod_{\text{inner}} P\left( S_{l,i}^M\right)\left( 1-P\left( S_{l,i}^M\right)\right),
\end{equation}
and the alignment likelihood
\begin{equation}
L^{||}_{l,\mathbf{D}}(M) := 4^2 \prod_{||} P\left( S_{l,i}^M\right)\left( 1-P\left( S_{l,i}^M\right)\right) .
\end{equation}
The inner likelihood measures anomalies inside a given multipole, while the outer likelihood measures the anomalies with respect to some outer direction. Eventually, the alignment likelihood measures the combined effect of alignment with an outer direction and intramultipole alignment.

\begin{figure}
\resizebox{\hsize}{!}{\includegraphics{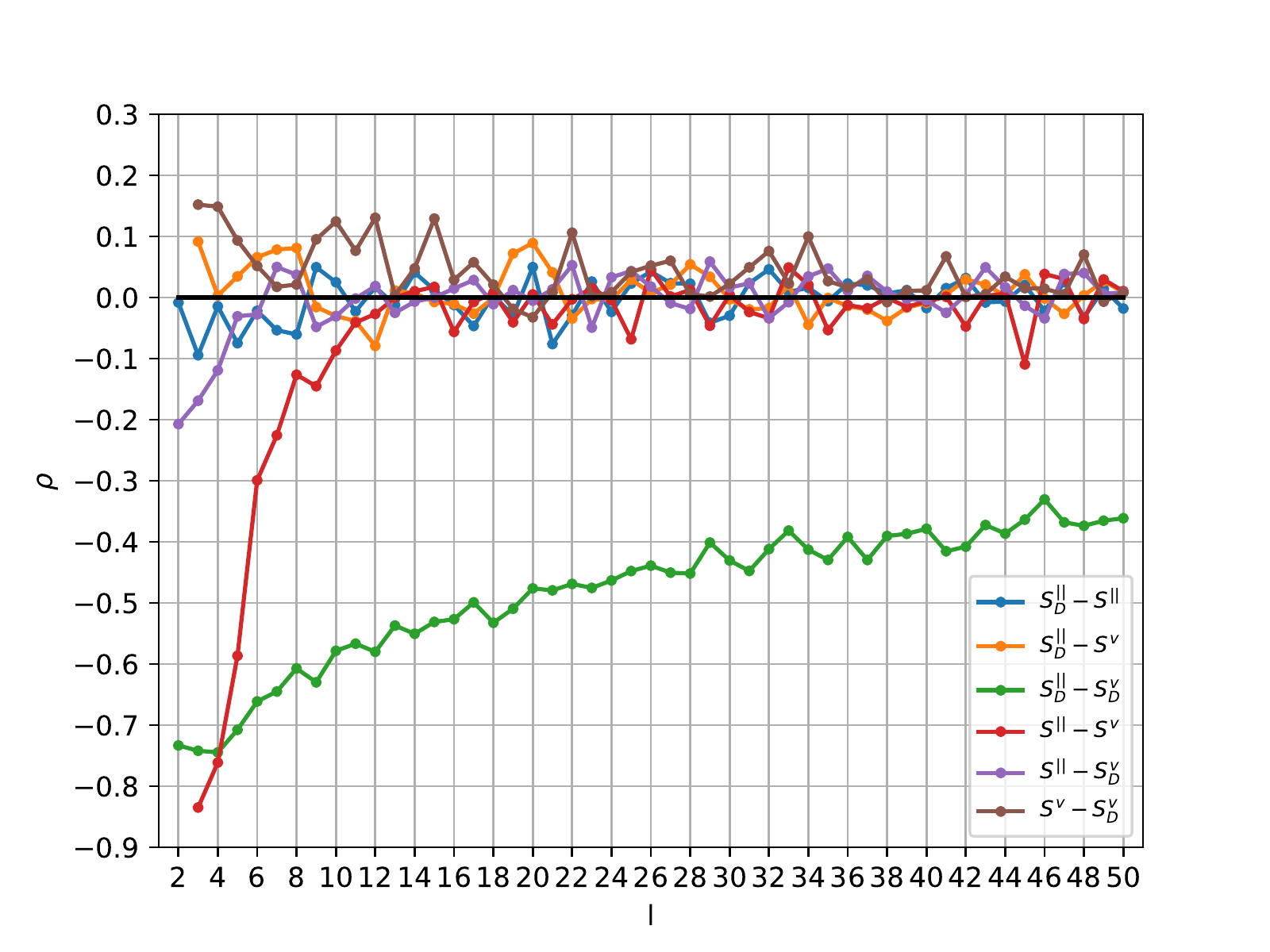}}
\caption{Linear correlation coefficients of statistics based on a Monte Carlo simulation with 1000 ensembles of $a_{lm}$. The combination $S^v-S^v_{\mathbf{D}}$ is the one mainly used in previous studies by means of correlations between area vectors. It shows only slight correlation at the largest angular scales.}
\label{lincorr}
\end{figure}

Before studying the Planck maps we need to understand the correlation between statistics. For low $l$ Fig.~\ref{lincorr} shows that the outer as well as the inner statistics are highly (anti)correlated. For higher $l$ the linear correlation of inner statistics vanishes, the outer statistics keep their correlation on a wide range of scales. Apart from statistical fluctuations the alignment statistics show nearly no correlation at all. Naively one would expect the inner and outer correlations to effect the behavior of the likelihoods, but the distribution of likelihoods in the range $2 \leq l \leq 50$ is nearly the same for all three considered types of likelihoods.

\subsection{Simulations}

We generate a fixed set of $1000$ ensembles of Gaussian and isotropic random $a_{lm}$ on the range $2 \leq l \leq 50$ while assuming Planck 2015 best-fit cosmological data to be fixed. Then we use the MPV calculation program \cite{CopiProg} to extract MPVs for each of the $1000$ ensembles and for the four full-sky foreground cleaned Planck 2015 CMB maps. From the MPVs we calculate the statistics described before. Then, using the fixed set of $1000$ random ensembles we calculate mean values, confidence levels, p-values and likelihoods and use them for analysis of all four Planck maps.

\section{Test directions}
\label{directions}

We use the following four physical directions, whose possible influences should have different and independent physical reasons:
\begin{itemize}
\item The cosmic dipole $(l,b) = (\SI{264.00}{\degree},\SI{48.24}{\degree})$ (with 
an amplitude $(\delta T /T_0)_{\text{dip}} = (3.3645\pm 0.002) \times 10^{-3}$), taken from \cite{Planck2015I}. The CMB dipole is assumed to be due to the peculiar motion of the Solar System with respect to the cosmic comoving frame \cite{Peebles1968}.
A correlation with this direction could imply that the nature of the kinematic dipole is not fully understood yet, that it has not been removed from the data properly, that the CMB contains an intrinsic dipole for itself, or that the calibration pipeline is odd.
\item The Galactic Pole $(l,b) = (\SI{0}{\degree},\SI{90}{\degree})$. Galactic foregrounds which are aligned with the disk of the Milky Way could give rise to an alignment with the Galactic Pole.
\item The Galactic Center $(l,b) = (\SI{0}{\degree},\SI{0}{\degree})$. The foreground pollution due to the inner part of the Milky Way could still be present in the cleaned maps. A correlation with this direction would indicate that these residuals still play an important role in data analysis.
\item The ecliptic pole $(l,b) = (\SI{96.38}{\degree},\SI{29.81}{\degree})$ (transformed from ecliptic to Galactic coordinates with the NASA conversion tool \footnote{\url{https://lambda.gsfc.nasa.gov/toolbox/tb\_{}coordconv.cfm}}). The lowest multipoles are known to correlate unusually with the ecliptic. Foreground pollution from the Solar System could cause such a correlation.
\end{itemize}

\begin{figure}
\includegraphics[width = \hsize]{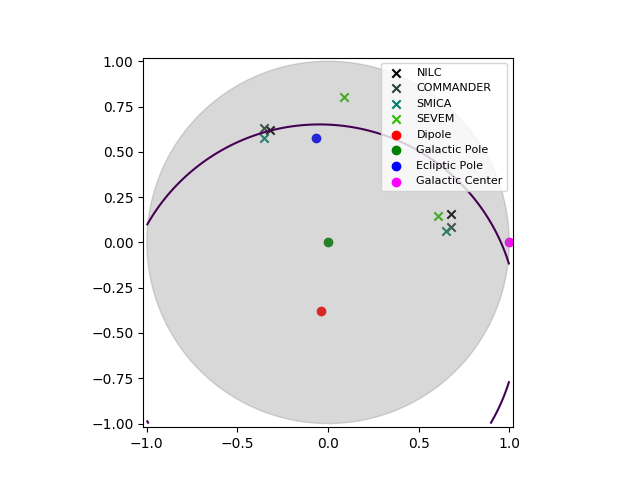}
\caption{MPVs for $l=2$ and physical directions in stereographic projection. The violet curve shows the plane orthogonal to the cosmic dipole.}
\label{stereo_2}
\end{figure}

In Fig.~\ref{stereo_2} we plot the MPVs for all pipelines at $l=2$ together with the four outer directions and the intersection of the plane orthogonal to the cosmic dipole with the celestial sphere in stereographic projection from the south pole. One should note that the stereographic projection does not preserve distances. Arcs close to the south pole get stretched with respect to arcs close to the north pole. But since we only consider one hemisphere, distances of points on the sphere are approximately conserved. Despite this disadvantage the stereographic projection was chosen because it allows for a simple and straightforward interpretation and has a nice geometrical meaning. Note that due to the identification of antipodal MPVs, opposite points on the unit circle in stereographic projection have to be identified. For $l=2$ the plot already shows one feature that we will encounter in Sec.~\ref{results}, namely that the MPVs nearly lie in the plane orthogonal to the cosmic dipole.

In Appendix \ref{app_stereo} further stereographic projection plots for $l=3,4$ (Fig.~\ref{stereo_low}) and $l=48,49$ (Fig.~\ref{stereo_high}) can be found. $l=2,3,4$ are plotted because at these multipoles the most anomalous behavior can be observed and $l=48,49$ have been chosen because they depict two higher multipoles which are orthogonal in the sense that one is especially unlikely and one is especially normal.

\section{Results}
\label{results}

The four statistics mentioned in Sec.~\ref{Analysis} were calculated for all four full-sky maps using all test directions and compared to the statistics of one thousand Monte Carlo ensembles of Gaussian and isotropic $a_{lm}$. We provide a qualitative description of the results in Secs.~\ref{sectdip}-\ref{sectinner} before before summarizing the findings in more precise statistical statements in Sec.~\ref{sectmultip}.

\subsection{Reproduction of known large scale anomalies and investigation of intermediate scales with outer vertical statistic}
\label{repro}
\begin{figure}
\begin{subfigure}[c]{\hsize}
\includegraphics[width = \hsize]{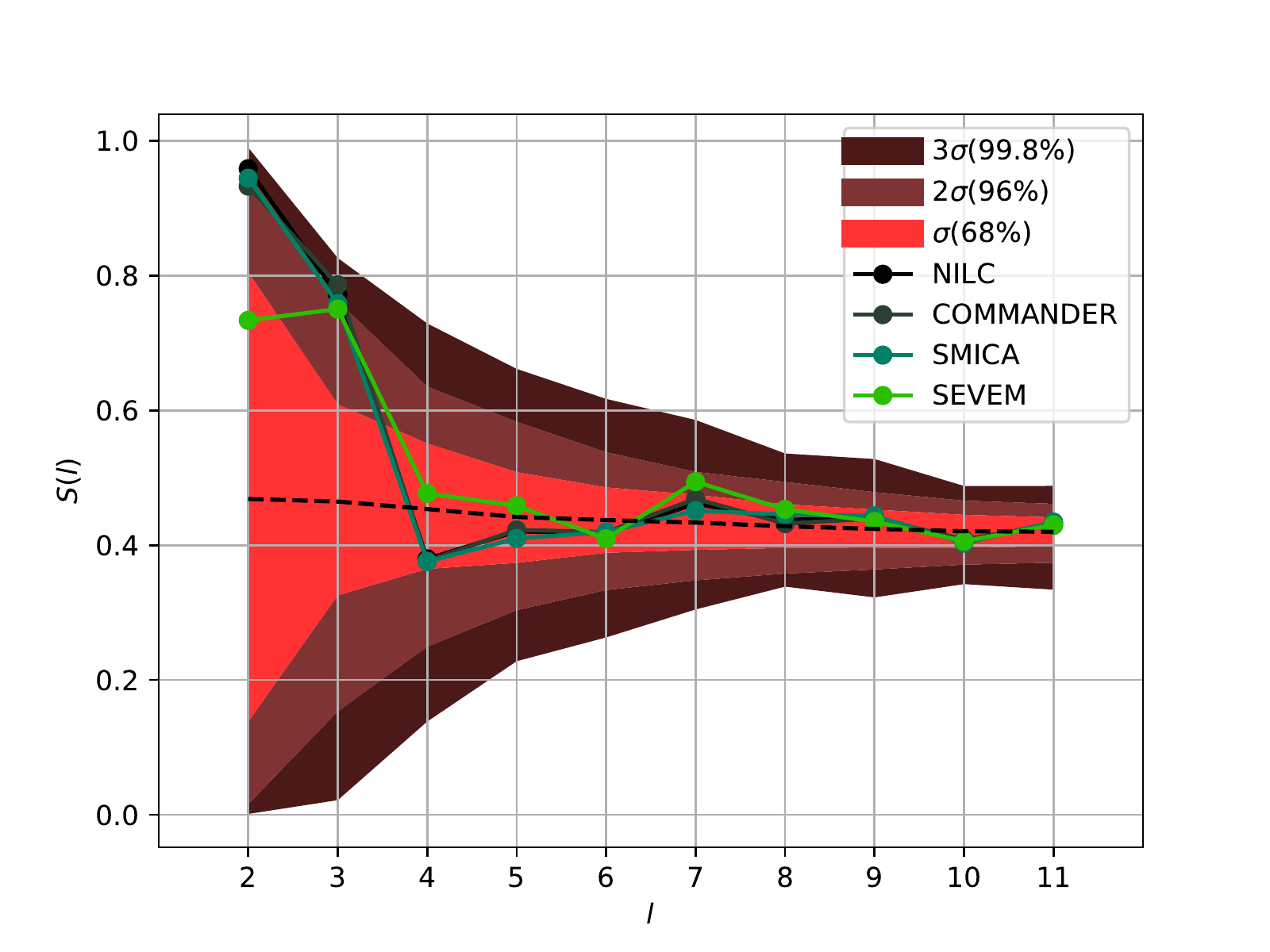}
\subcaption{Large angular scales $2\leq l \leq 11$}
\label{S5_dip_small}
\end{subfigure}
\begin{subfigure}[c]{\hsize}
\includegraphics[width = \hsize]{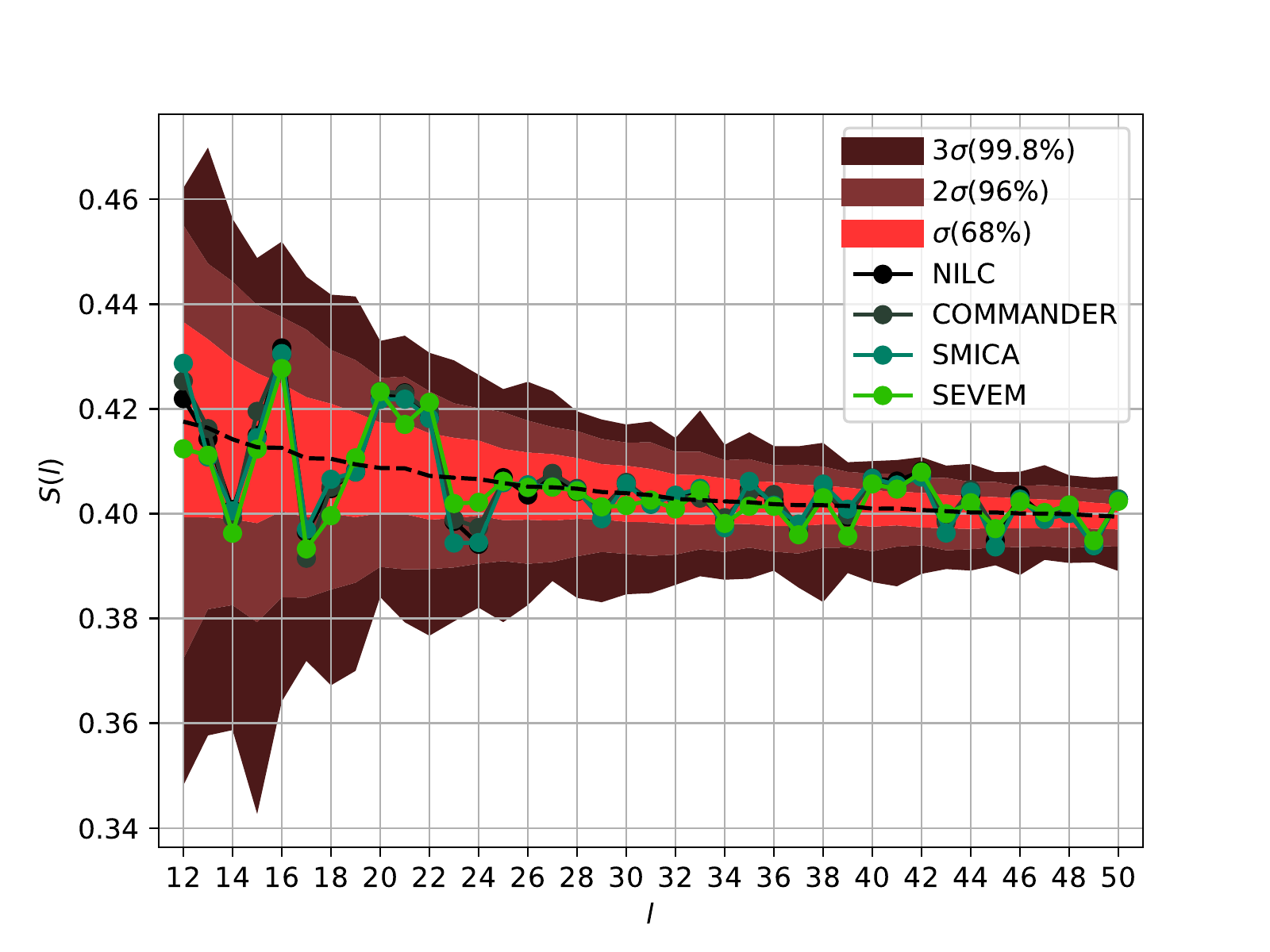}
\subcaption{Smaller angular scales $12 \leq l \leq 50$}
\label{S5_dip_large}
\end{subfigure}
\caption{Comparison of pipelines for $S^v_{\mathbf{D}}$ with $\mathbf{D}$ the cosmic Dipole. The expectation value and $1,2,3\sigma$ regions from Monte Carlo simulations are included.}
\label{S5_dip}
\end{figure}

It has been observed in previous studies that on the largest angular scales, i.e. the quadrupole and octupole, the MPVs correlate with the cosmic dipole. 
Fig.~\ref{S5_dip_small} shows that this correlation is due to a strong orthogonality of the MPVs and the dipole direction. The area vectors of the quadrupole and 
octupole in COMMANDER, NILC and SMICA show an alignment with the dipole at $2\sigma$ level. By visualizing the quadrupole as a plane, this means that the 
cosmic dipole direction is nearly perfectly orthogonal to this plane which cannot be achieved in about 96$\%$ of random ensembles of Gaussian and isotropic 
temperature fluctuation fields. The quadrupole value of SEVEM seems less anomalous, but as we will argue later, we find hints that SEVEM still shows residual foreground 
effects via a correlation of the MPVs with the Galactic Center and especially with the Galactic Pole. That the MPVs of the quadrupole are almost normal to the dipole direction is also shown in Fig.~\ref{stereo_2}.

Fig.~\ref{S5_dip_large} shows the vertical outer statistic for smaller angular scales. The region $20 \leq l \leq 24$ sticks out just like the largest angular scales. 
At these scales data points outside the $1\sigma$ regions cluster. Both scale ranges show a similar behavior; first the MPVs are too close to the plane orthogonal to the cosmic dipole, and then they are too far away from this plane. It should be noted that the two suspicious scales (large and intermediate) coincide with the scales at which the measured angular power spectrum deviates from the best-fit $\Lambda$CDM model of the Planck 2015 analysis \cite{Planck2015I}. This hints towards a connection between the power spectrum deviation and the peculiar motion of the Solar System with respect to the cosmic frame.

\subsection{Comparison of directions and pipelines using aligned statistics}

In Figs.~\ref{S1_dip}--\ref{S1_ep} we plot the outer statistic $S^{||}_{\mathbf{D}}$ for each of the five directions including the $1\sigma$ to $3\sigma$ regions from the Monte Carlo simulations in the range of large angular scales $2 \leq l \leq 11$ and in the range of smaller angular scales $12 \leq l \leq 50$, comparing in each plot all four pipelines. Figure \ref{S3} shows the same for the inner statistic $S^{||}$. Figure \ref{LH_outer_dip} shows the outer likelihood as a function of $l$ for the cosmic dipole. In Table \ref{multinomial} we present multinomial probabilities and respective p-values.
\subsubsection{Cosmic dipole}
\label{sectdip}
\begin{figure}
\begin{subfigure}[c]{\hsize}
\includegraphics[width = \hsize]{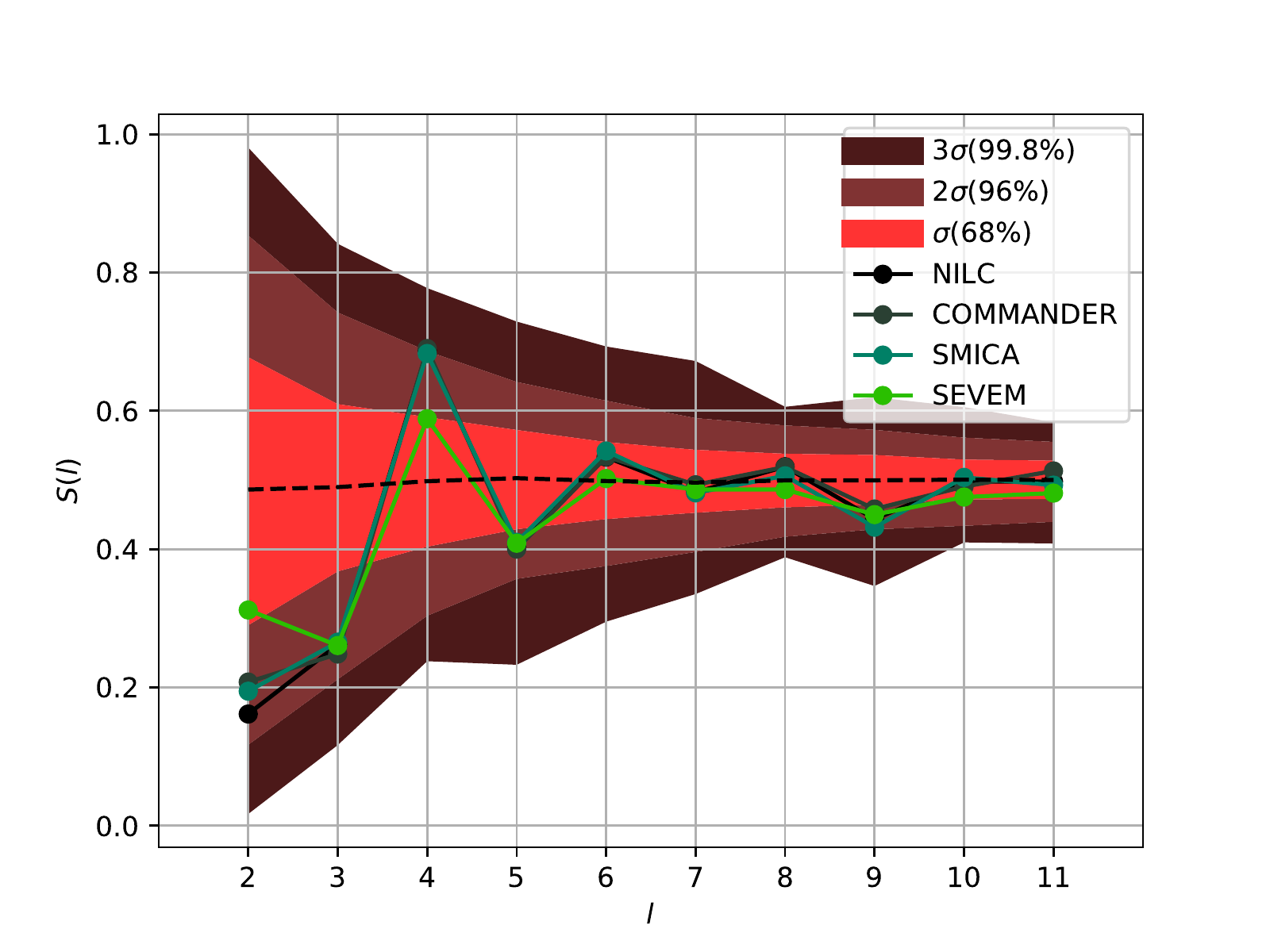}
\subcaption{Large angular scales/small multipoles}
\label{S1_dip_small}
\end{subfigure}
\begin{subfigure}[c]{\hsize}
\includegraphics[width = \hsize]{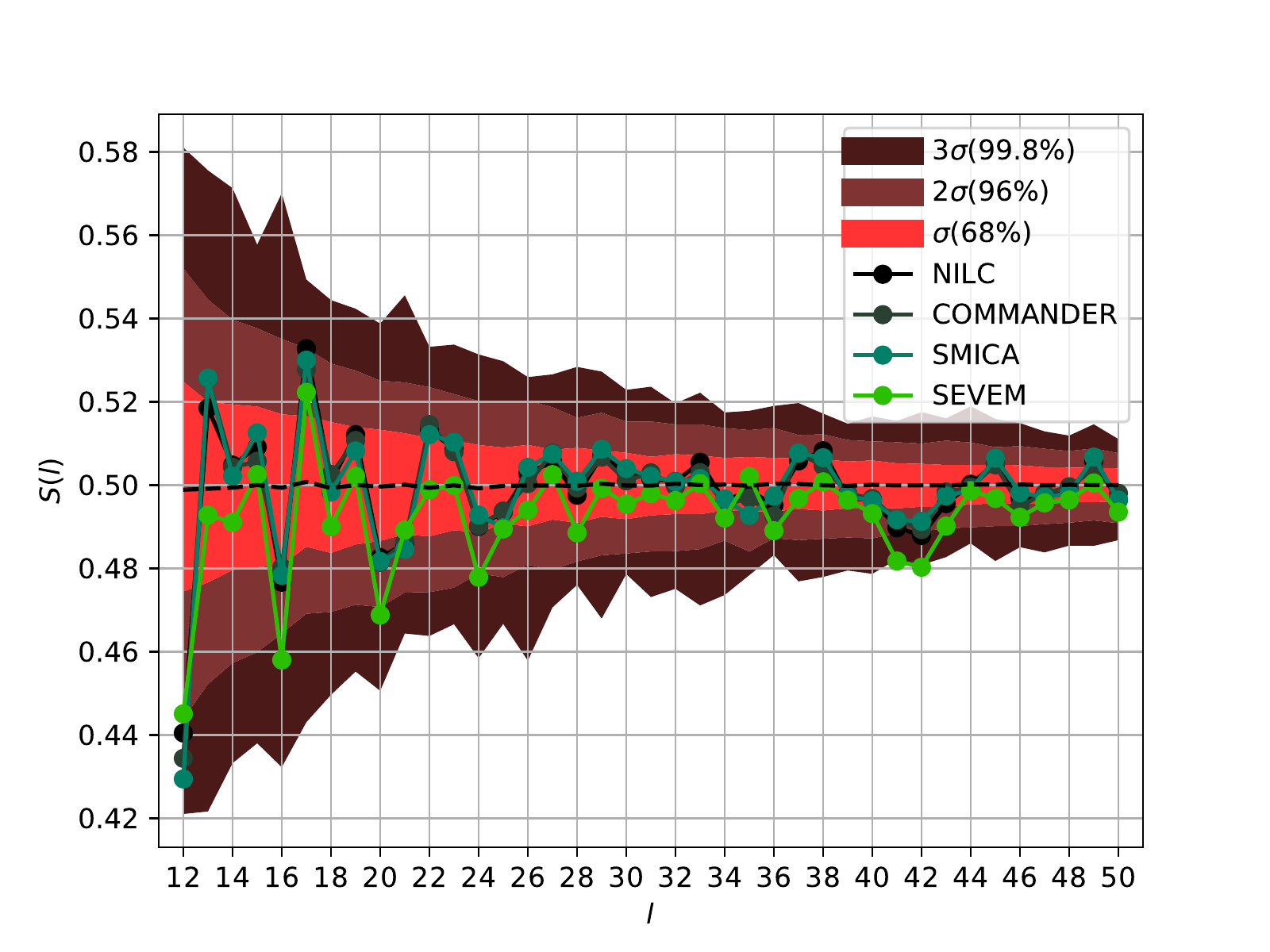}
\subcaption{Small angular scales/large multipoles}
\label{S1_dip_large}
\end{subfigure}
\caption{Comparison of pipelines for $S^{||}_{\mathbf{D}}$ with $\mathbf{D}$ the cosmic dipole. The expectation value and $1,2,3\sigma$ regions from Monte Carlo simulations are included.}
\label{S1_dip}
\end{figure}

One observes (see Fig.~\ref{S1_dip_small}) that the large scale anticorrelation of the quadrupole and octupole with the cosmic dipole -- see for example the review \cite{Schwarz2015} -- is still present in the second release data. While for SEVEM the antialignment is more pronounced at $l=3$ than at $l=2$, both multipole data points are equally unusual in the other three pipelines (both nearly $2\sigma$). It turns out that for $l=4$ an even less expected alignment of the COMMANDER, NILC and SMICA data with the dipole is present. Except for SEVEM each of the lowest multipoles $l=2,3,4,5$ shows an unexpected behavior with respect to the cosmic dipole.

The large multipole behavior (see Fig.~\ref{S1_dip_large}) already shows a clear deviation of SEVEM from the other three pipes. On the whole range $12 \leq l \leq 50$ SEVEM is less aligned with the cosmic dipole than the other cleaned maps and it admits more unlikely data points. 

Concerning COMMANDER, SMICA and NILC, all in all 17 out 49 multipoles are outside of the $1\sigma$ region. Despite the large angular scale $2 \leq 5$ there is no other clustering of at least four unlikely multipoles in a row. The conclusion here is that for $l \geq 5$ the data follow the statistically expected behavior. Furthermore neither alignment nor antialignment is preferred. 

Eventually we state that COMMANDER, NILC and SMICA show a very similar behavior and deviate less from each other than one would naively expect. It seems that for this type of data analysis the cleaning algorithms (except for SEVEM) all have the same quality on the considered range of scales and that the precise choice of the cleaning algorithm does not affect our results. Since the three algorithms use different frequency bands and different masks the strong coincidence surprises.

\subsubsection{Galactic Pole}

\begin{figure}
\begin{subfigure}[c]{\hsize}
\includegraphics[width = \hsize]{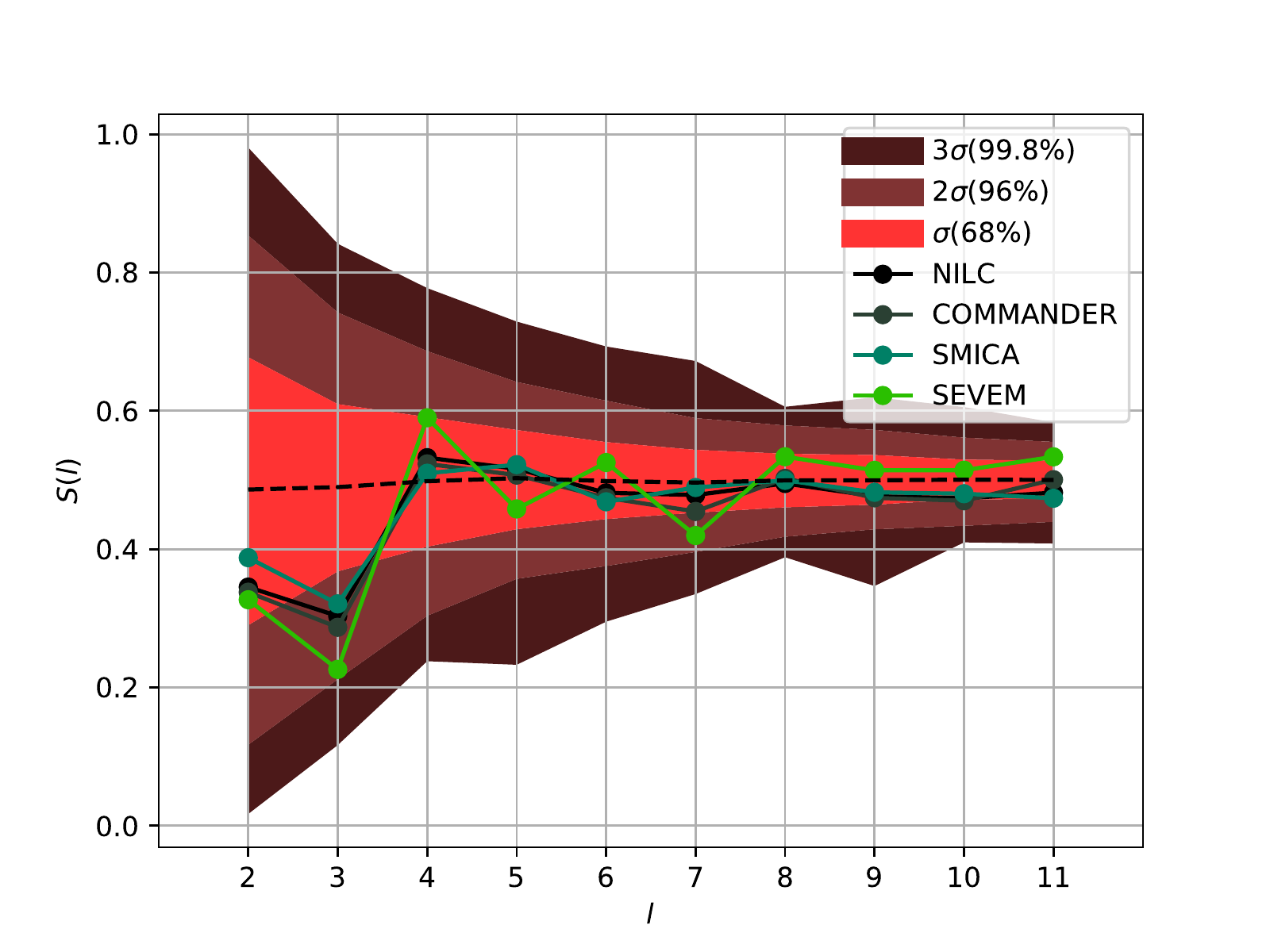}
\subcaption{Large angular scales/small multipoles}
\label{S1_gp_small}
\end{subfigure}
\begin{subfigure}[c]{\hsize}
\includegraphics[width = \hsize]{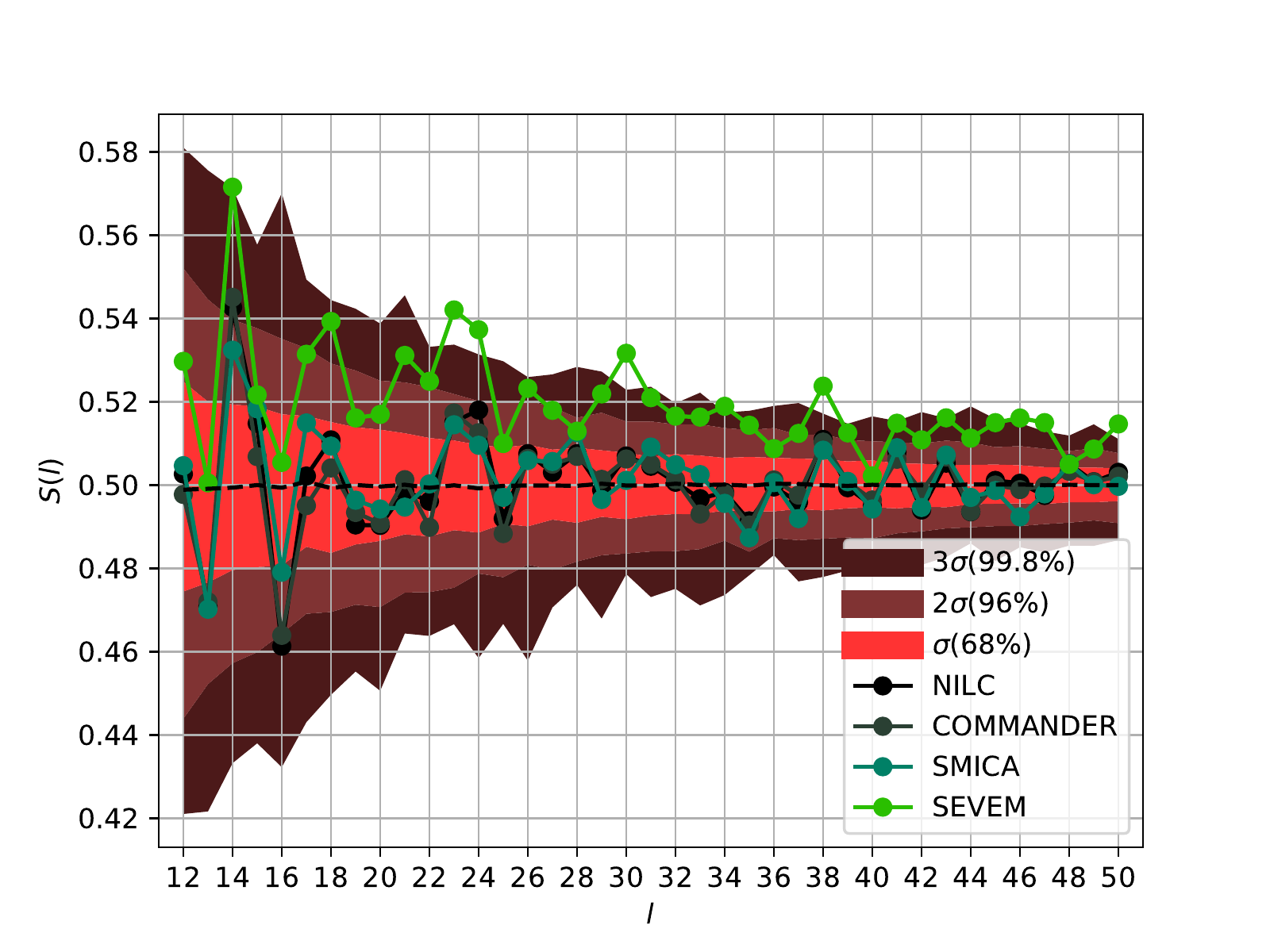}
\subcaption{Small angular scales/large multipoles}
\label{S1_gp_large}
\end{subfigure}
\caption{Comparison of pipelines for $S^{||}_{\mathbf{D}}$ with $\mathbf{D}$ the Galactic Pole. The expectation value and $1,2,3\sigma$ regions from Monte Carlo simulations are included.}
\label{S1_gp}
\end{figure}

The large scale behavior with respect to the Galactic Pole is as expected 
(even in the SEVEM map); see Fig.~\ref{S1_gp_small}. When referring to larger multipoles (see Fig.~\ref{S1_gp_large}) SEVEM shows even stronger deviations from the other maps than in the other directions. From $l=12$ on SEVEM is tremendously aligned with the Galactic Pole. Eight out of 39 data points lie even outside of the $3\sigma$ region. Again, we state that this behavior might be a hint towards Milky Way residuals in the SEVEM map.

When comparing Fig.~\ref{S1_gp} to Fig.~\ref{With_mask_gp} from Appendix \ref{app_mask}, where the statistic $S^{||}_{\mathbf{D}}$ is plotted for the Galactic Pole for all four pipelines but with the SEVEM mask applied to the map, it becomes obvious that the strong alignment of SEVEM with the Galactic Pole is solely due to the masked region. When the mask is applied, all four maps show a similar behavior and the deviation of SEVEM from the others vanishes nearly completely, especially in the high $l$ regime. This is not surprising since it is assumed by the Planck Collaboration itself that SEVEM carries residual effects of the Galactic Plane.

Hence, one concludes that one should be careful when using SEVEM for full-sky analyses. A more detailed investigation of the precise cleaning algorithm needs to be taken into account.

Finding such a strong deviation of the SEVEM map from complete
randomness is a confirmation of the power of MPV to identify alignment effects.

Concerning the other maps neither alignment nor antialignment is preferred.

Altogether, on the whole range $2 \leq l \leq 50$ the Galactic Pole incorporates more low probability multipoles than the Galactic Center or the ecliptic pole but approximately as many as the cosmic dipole. 

\subsubsection{Galactic Center}

\begin{figure}
\begin{subfigure}[c]{\hsize}
\includegraphics[width = \hsize]{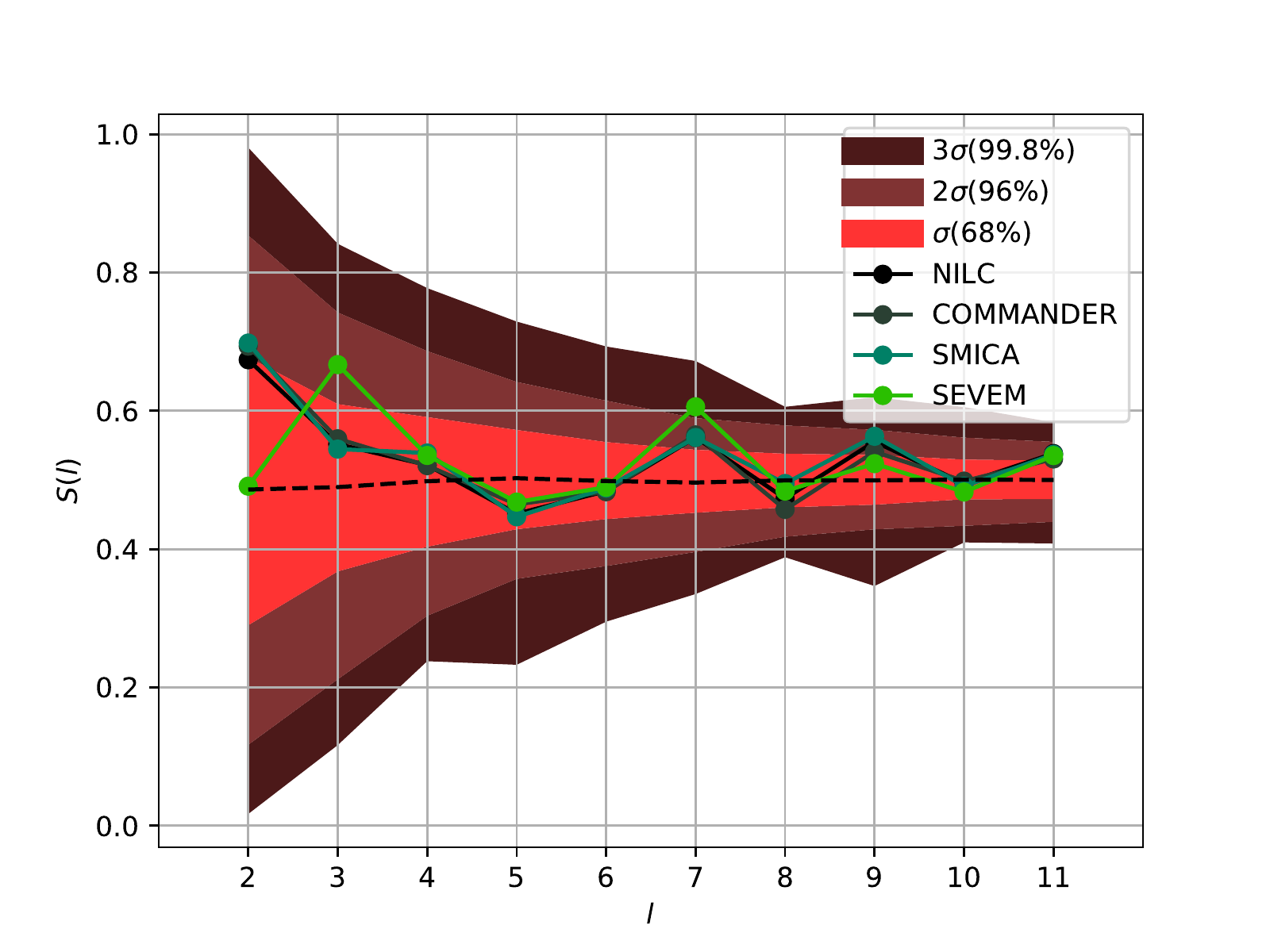}
\subcaption{Large angular scales/small multipoles}
\label{S1_gc_small}
\end{subfigure}
\begin{subfigure}[c]{\hsize}
\includegraphics[width = \hsize]{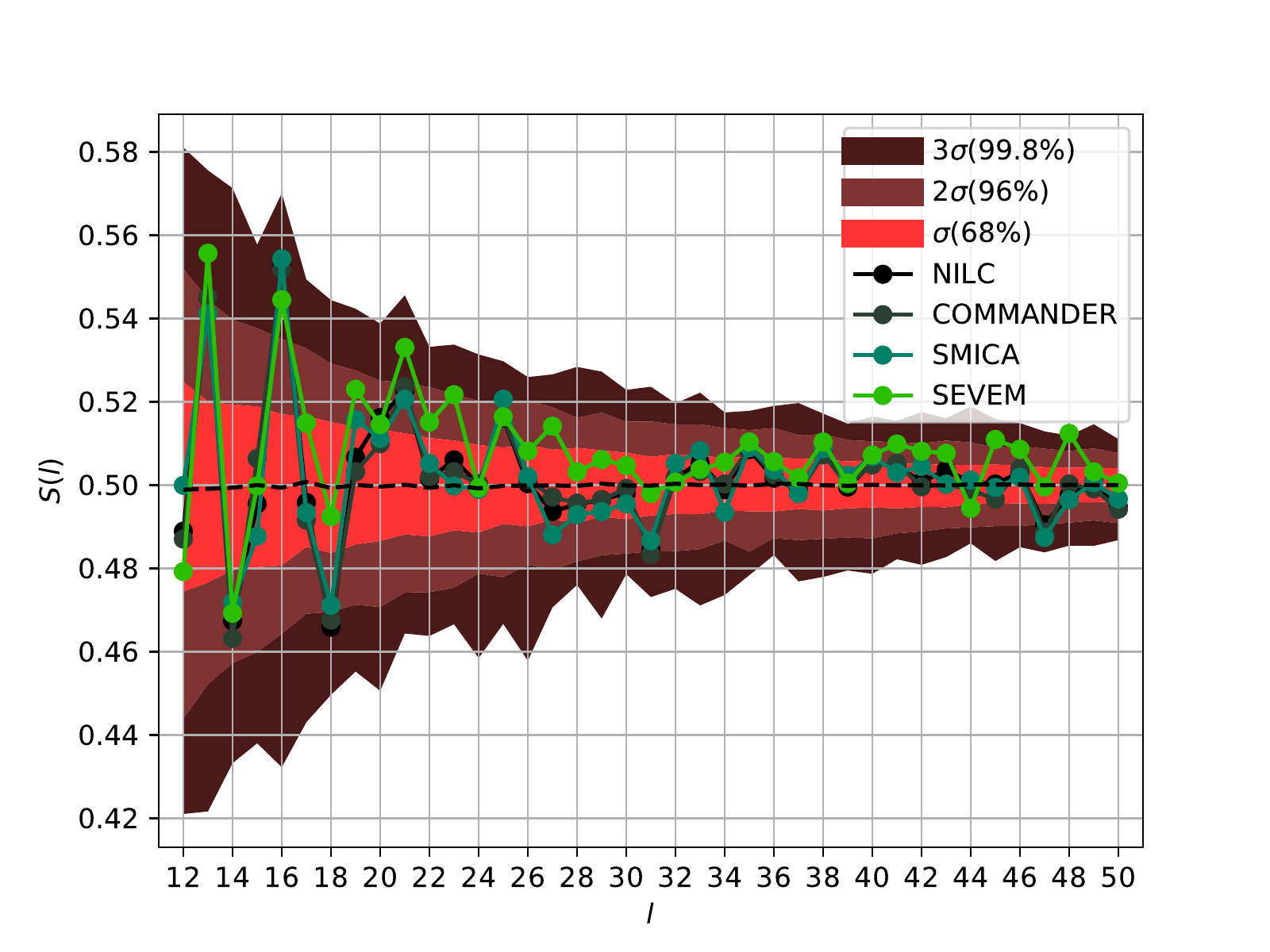}
\subcaption{Small angular scales/large multipoles}
\label{S1_gc_large}
\end{subfigure}
\caption{Comparison of pipelines for $S^{||}_{\mathbf{D}}$ with $\mathbf{D}$ the Galactic Center. The xpectation value and $1,2,3\sigma$ regions from Monte Carlo simulations are included.}
\label{S1_gc}
\end{figure}

The behavior with respect to the Galactic Center tends to be less unexpected than the cosmic dipole on large angular scales; see Fig.~\ref{S1_gc_small}. Only $l=2,7$ and $9$ lie just outside of $1\sigma$. Again SEVEM deviates clearly from the other maps on smaller angular scales (see Fig.~\ref{S1_gc_large}), this time showing a stronger alignment with the Galactic Center, especially on the midrange scales which correspond approximately to the angular size of the Galactic core.

There are three multipoles far away from the expectation in the non-SEVEM maps. The multipoles $l=16,18$ and $47$ have data points outside of the $2\sigma$ regions 
and $l=16$ is nearly at $3\sigma$.

\subsubsection{Ecliptic pole}

\begin{figure}
\begin{subfigure}[c]{\hsize}
\includegraphics[width = \hsize]{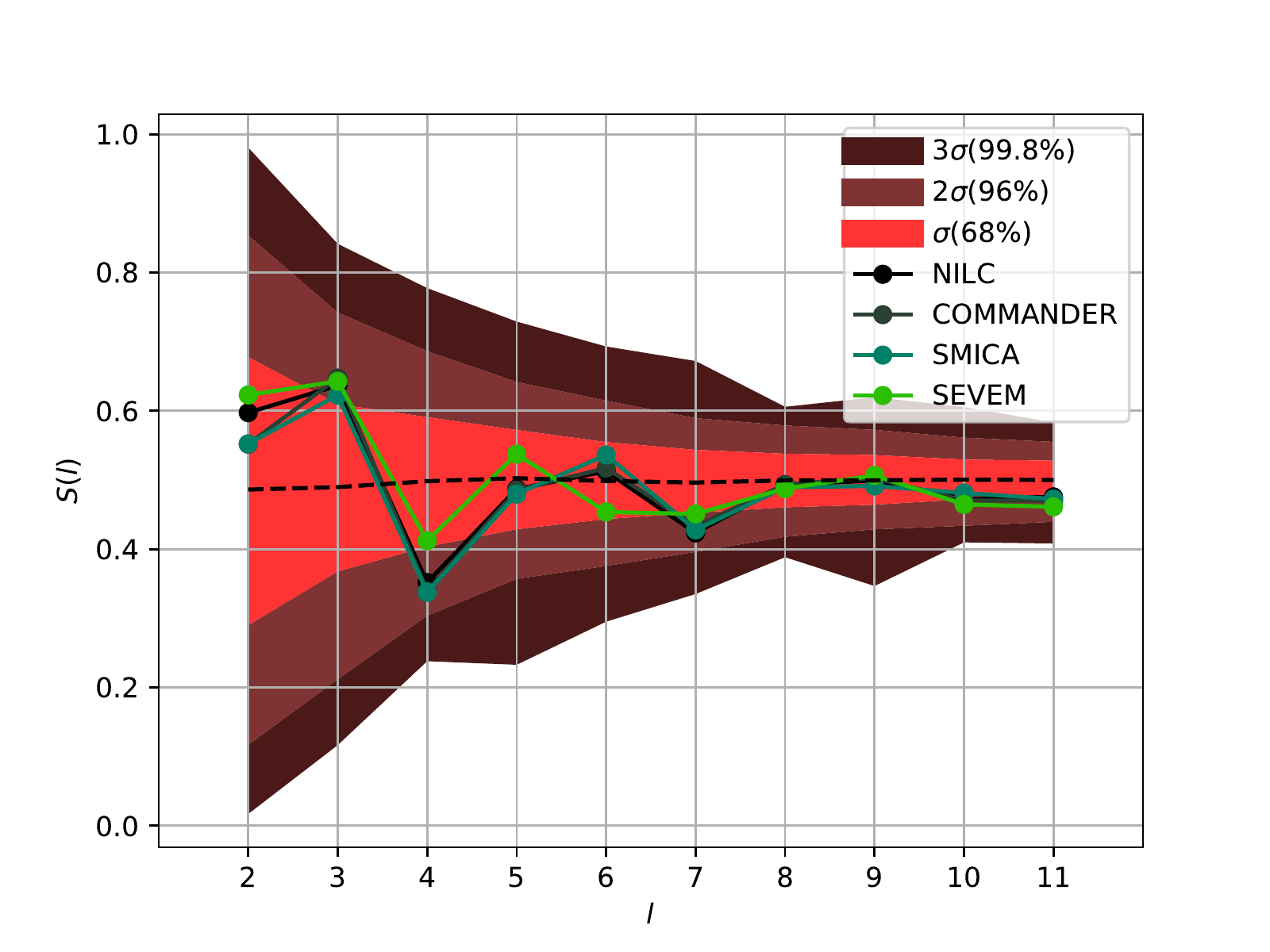}
\subcaption{Large angular scales/small multipoles}
\label{S1_ep_small}
\end{subfigure}
\begin{subfigure}[c]{\hsize}
\includegraphics[width = \hsize]{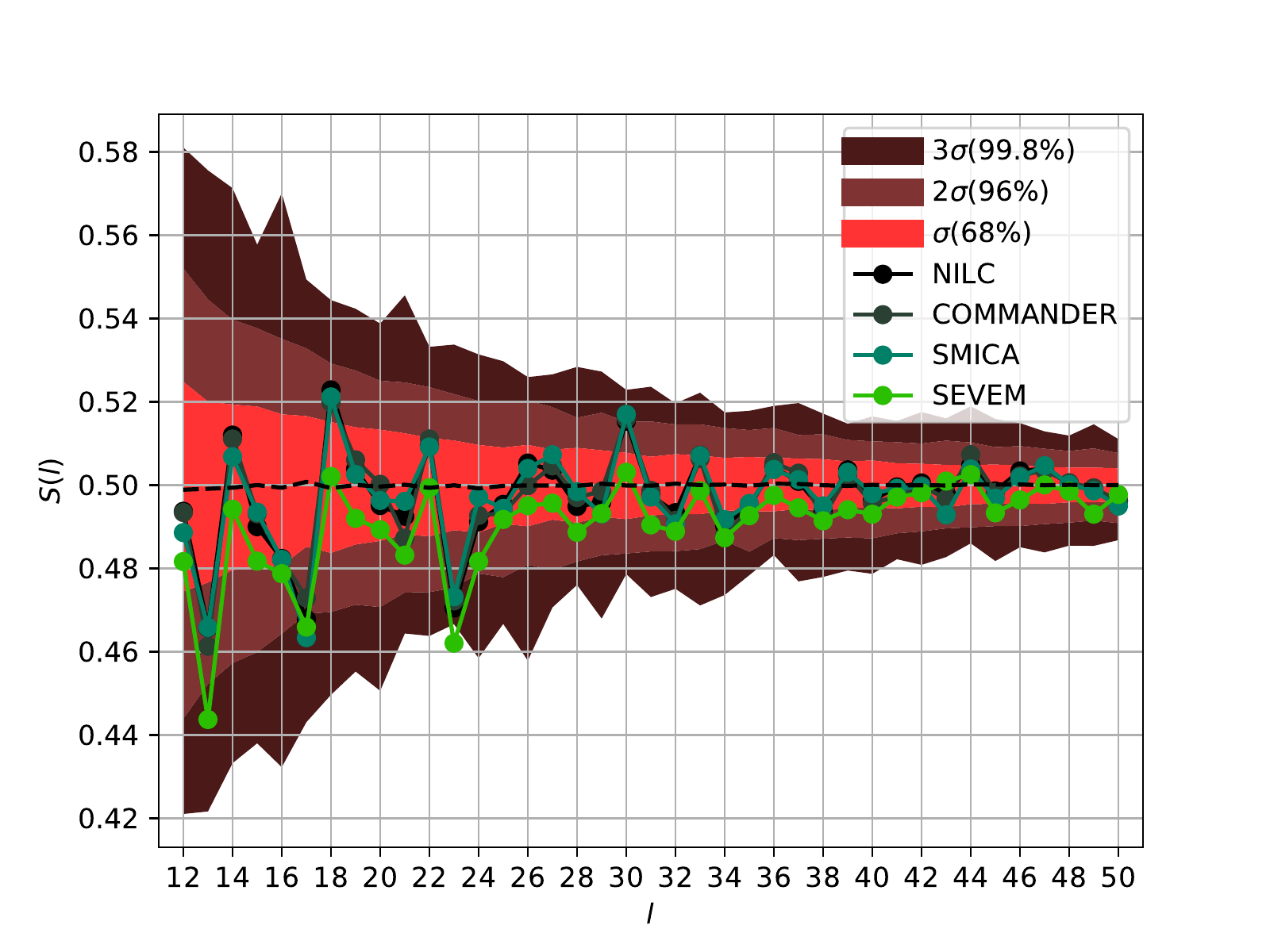}
\subcaption{Small angular scales/large multipoles}
\label{S1_ep_large}
\end{subfigure}
\caption{Comparison of pipelines for $S^{||}_{\mathbf{D}}$ with $\mathbf{D}$ the ecliptic pole. The expectation value and $1,2,3\sigma$ regions from Monte Carlo simulations are included.}
\label{S1_ep}
\end{figure}

On large angular scales (see Fig.~\ref{S1_ep_small}), the data show an even more expected behavior with respect to the ecliptic pole than with respect to the cosmic dipole.

On smaller angular scales (see Fig.~\ref{S1_ep_large}) SEVEM again clearly deviates from the other maps, showing more antialignment with the ecliptic pole. For the other maps, the ecliptic seems to show less correlation with the CMB data than the dipole.

\subsubsection{Inner alignment}
\label{sectinner}
\begin{figure}
\begin{subfigure}[c]{\hsize}
\includegraphics[width = \hsize]{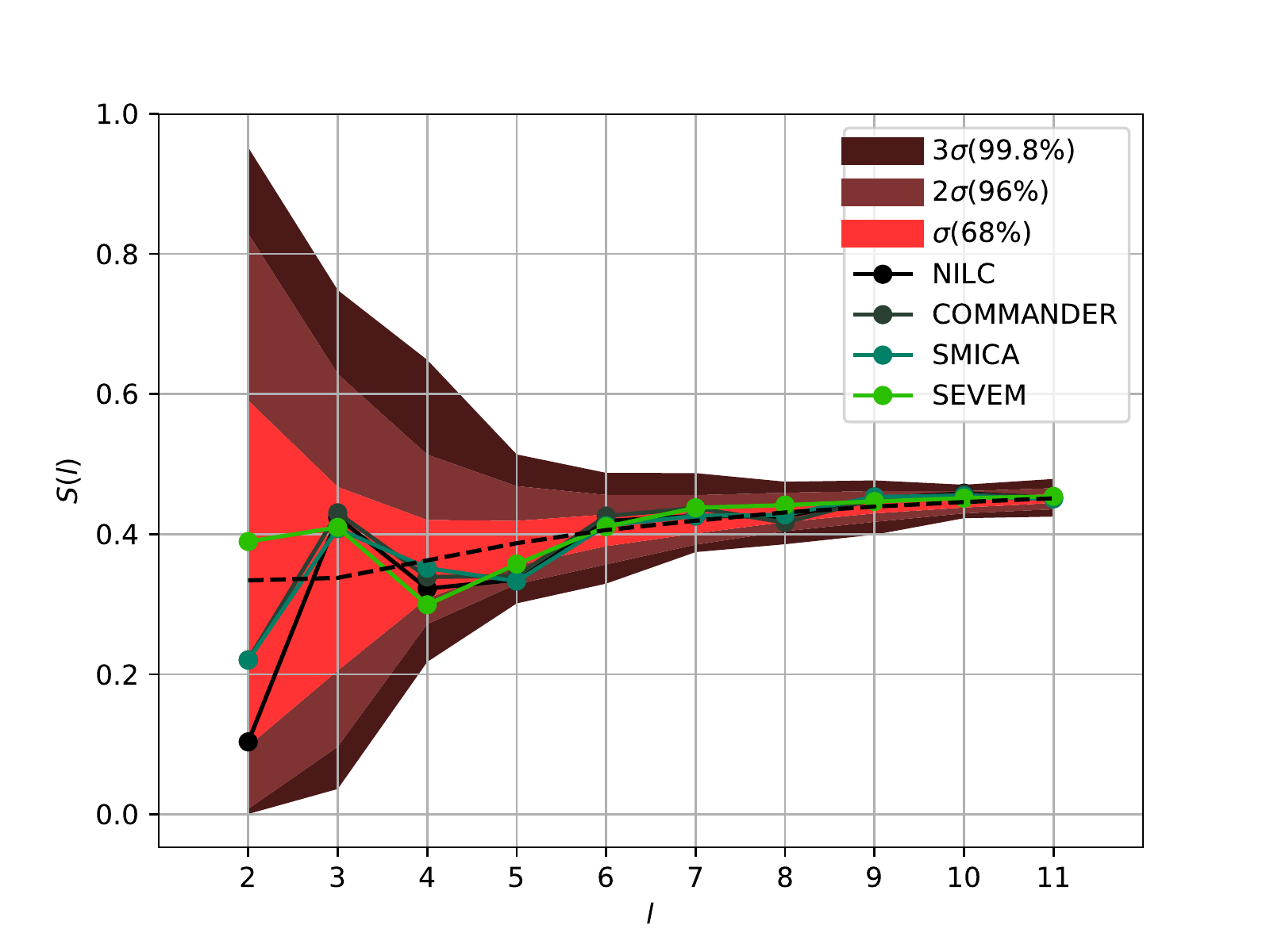}
\subcaption{Large angular scales/small multipoles}
\label{S3_small}
\end{subfigure}
\begin{subfigure}[c]{\hsize}
\includegraphics[width = \hsize]{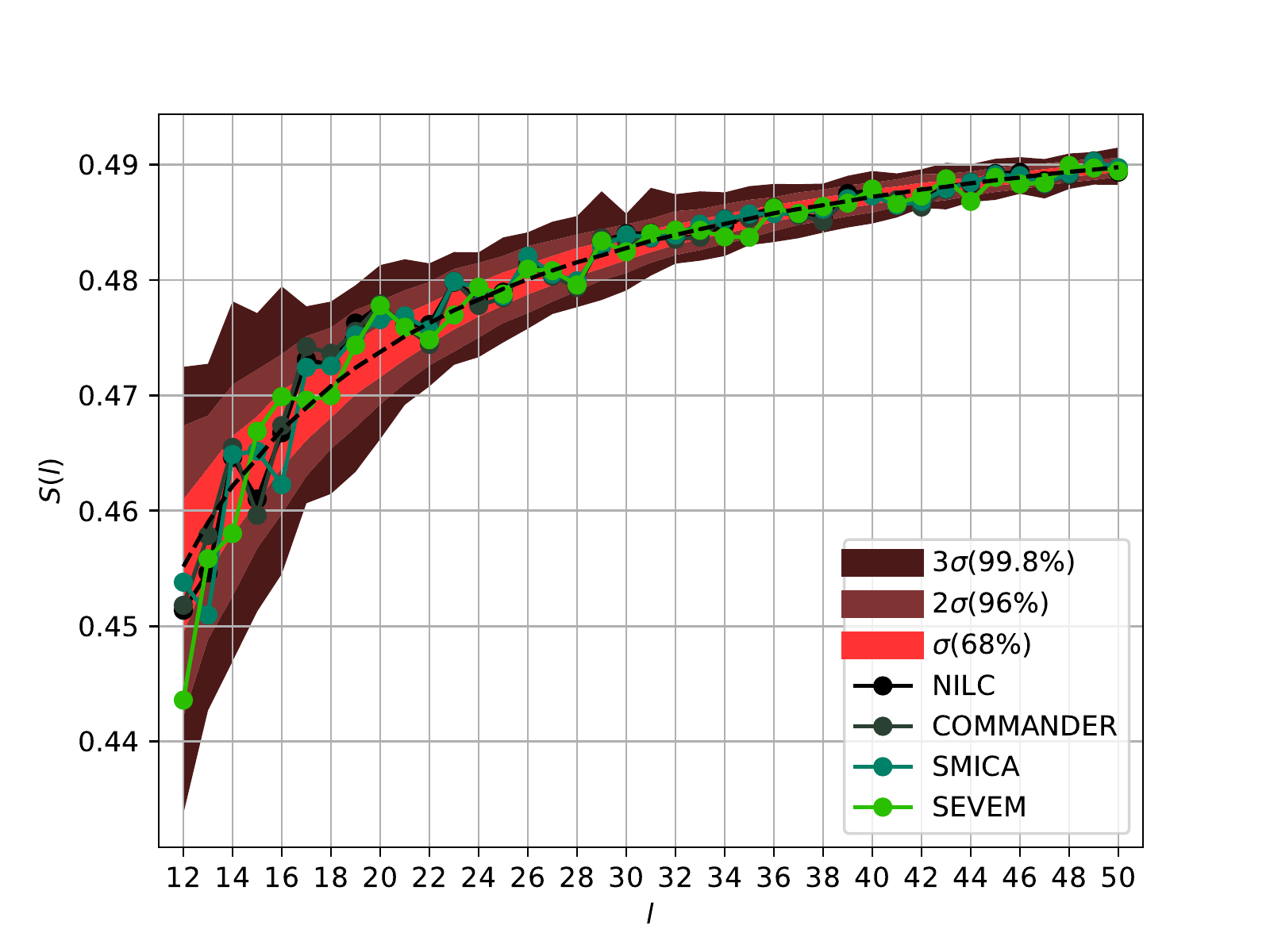}
\subcaption{Small angular scales/large multipoles}
\label{S3_large}
\end{subfigure}
\caption{Comparison of pipelines for $S^{||}$. The expectation value and $1,2,3\sigma$ regions from Monte Carlo simulations are included.}
\label{S3}
\end{figure}

The inner statistic $S^{||}$ comes equipped with a much smaller variance than the outer statistic $S^{||}_{\mathbf{D}}$, as can be seen in Fig.~\ref{S3}. Hence, the inner statistic is more susceptible to computational errors or minor fluctuations than the outer statistic, but nevertheless the plots paint a distinct picture: the inner statistic of COMMANDER, NILC and SMICA lies inside $1\sigma$ for most of the multipoles and does not leave $2\sigma$ one single time. No strong outliers are present, either in the low $l$ or in the large $l$ regime. If anomalies are present in the CMB, they seem to be mainly caused by correlation with outer directions, while no remarkable intramultipole correlation of MPVs can be observed. Note that other methods than ours could reveal hidden intramultipole correlations that cannot be observed with our simple method.

\subsubsection{Multinomial p-values}
\label{sectmultip}

In Table \ref{multinomial} we gather the number of multipoles in the range 
$2\leq l \leq 50$ lying within $1\sigma$ ($n_1$), between $1\sigma$ and $2\sigma$ ($n_2$) as well as outside of $2\sigma$ ($n_3$) and the probability of that based on a multinomial distribution 
\begin{equation}
\mathfrak{p}(n_1, n_2, n_3) = \binom{49}{(n_1,n_2,n_3)}(0.68)^{n_1}(0.28)^{n_2}(0.04)^{n_3}
\end{equation} 
for the statistics $S^{||}_{\mathbf{D}}$ and $S^{||}$. We also define a corresponding multinomial p-value, which is the probability to find at least $n_2$ and $n_3$ multipoles at $1\sigma$ to $2\sigma$ and above $2 \sigma$ deviation from expectation, i.e.
\begin{equation} 
\text{multinomial p-value} = \sum_{i = 0}^{n_1} \sum_{k = 0}^{i+n_2} 
\mathfrak{p}(n_1 - i, n_2 + i - k, n_3 + k).
\end{equation}
The smaller the p-value the more the data deviate from the expectation concerning the number of outliers. An alternative definition of the p-value as the sum over all multinomial probabilities that are smaller than the given probability would result in higher p-values, but would be inadequate for our purposes since it would involve configurations which are unlikely normal, as for example $(n_1,n_2,n_3) = (49,0,0)$ as well.

\begin{table*}
\begin{tabular}{|c|c|c|c|c|c|c|}
\hline
Direction & Map & $<1\sigma$ & $1$--$2\sigma$ & $>2\sigma$ & $\mathfrak{p}_{\mathrm{multi}} (\%)$  & Multinomial p-value (\%) \\
\hline \hline
Cosmic dipole & COMMANDER & 31 & 14 & 4 & 1.06 & 6.41 \\
 & NILC & 32 & 13 & 4 & 1.12  & 8.11 \\
 & SMICA & 28 & 19 & 2 & 0.84 & 5.64 \\
 & SEVEM & 32 & 12 & 5 & 0.42 & 3.16 \\
\hline
Galactic Pole & COMMANDER & 37 & 11 & 1 & 2.33 & 79.30 \\
 & NILC & 36 & 12 & 1 & 2.96 & 73.90 \\
 & SMICA & 33 & 16 & 0 & 1.42 & 51.46 \\
 & SEVEM & 10 & 11 & 28 & $10^{-26}$ & $10^{-26}$  \\
\hline
Galactic Center & COMMANDER & 32 & 14 & 3 & 2.25  & 16.77 \\
 & NILC & 34 & 13 & 2 & 3.48  & 41.87 \\
 & SMICA & 29 & 17 & 3 & 1.15 & 6.16 \\
 & SEVEM & 26 & 17 & 6 & 0.04 & 0.14 \\
\hline
Ecliptic pole & COMMANDER & 32 & 15 & 2 & 3.15 & 27.59 \\
 & NILC & 36 & 10 & 3 & 1.33  & 28.80 \\
 & SMICA & 34 & 12 & 3 & 2.15 & 24.10 \\
 & SEVEM & 29 & 17 & 3 & 1.15 & 6.16 \\
\hline
Inner alignment & COMMANDER & 30 & 16 & 3 & 1.58 & 9.19 \\
& NILC & 32 & 17 & 0 & 1.13 & 39.43 \\
& SMICA & 32 & 17 & 0 & 1.13 & 39.43 \\
& SEVEM & 32 & 15 & 2 & 3.15 & 27.59 \\
\hline
\end{tabular}
\caption{Number of multipoles lying inside the $1\sigma$ region, between the $1\sigma$ and $2\sigma$ boundaries and outside of the $2\sigma$ region for all maps and directions for statistic $S_{\mathbf{D}}^{||}$ as well as for statistic $S^{||}$. In the last two columns we give the multinomial probability for these distributions of multipoles amongst the $\sigma$ regions in percent up to two digits and the respective p-values.
}
\label{multinomial}
\end{table*}

Inspecting Table \ref{multinomial}, the first thing to note is that SEVEM is strongly (anti)correlated with the Galactic Pole, as it shows a p-value of $10^{-26}$ \%. The correlation with the 
Galactic Center is also significant with a p-value of $0.14$ \%. 
All non-SEVEM maps behave as expected with respect to the Galactic Pole and the ecliptic pole. Regarding the inner alignment a slight deviation of COMMANDER from the other maps can be observed. While NILC and SMICA do not possess any data point outside of $2\sigma$, COMMANDER contains some data points shifted from $1\sigma$ to beyond $2\sigma$. Concerning the Galactic Center NILC seems more normal than COMMANDER and especially SMICA. However, the deviations are not extremely large and taking into account the other two directions, the similarity of all non-SEVEM maps emerges again. The cosmic dipole is the only considered direction for which all maps show p-values below $10\%$. Next, we investigate from which ranges this slight correlation of data and dipole stems.

\begin{figure}
\resizebox{\hsize}{!}{\includegraphics{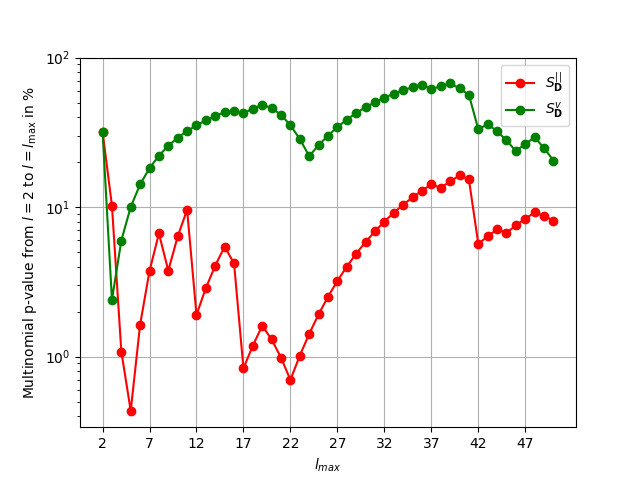}}
\caption{Multinomial p-values for outer statistics $S^{||}_{\mathbf{D}}$ and $S^{v}_{\mathbf{D}}$ with $\mathbf{D}$ the cosmic dipole. The p-value is calculated from $l=2$ to $l=l_{\mathrm{max}}$ using the NILC map.}
\label{multi_p_plot}
\end{figure}

The correlation of MPV with the cosmic dipole is studied in more detail in Fig.~\ref{multi_p_plot}. It 
shows the dependence of the multinomial p-value on $l_{\mathrm{max}}$ as one enlarges the 
considered multipole range from $[2,2]$ to $[2,l_{\mathrm{max}}]$. Here we additionally take into 
account statistic $S^{v}_{\mathbf{D}}$, which measures orthogonality of a multipole to the given 
direction, and only focus on the NILC map (it is most normal with respect to all the other tests 
considered here). We find that the correlation with the cosmic dipole is due to the largest 
angular scales and the region around $l=20$. The p-value of the aligned statistics drops to 
$0.4\%$ for the range $[2,5]$, while the p-value of the vertical statistic is below $2\%$ for $[2,3]$.
Around $l=20$ the p-value curve shows a dip with a minimum at $l=22$ and p-value of $0.7\%$ 
for the aligned statistic and a drop in the vertical statistic with a minimum at $l = 24$. 
A third region where the p-value of both statistics clearly drops is around $l=42$, but with 
higher multinomial p-value than at $l$ around $20$. It is remarkable, that these three regions 
are exactly those regions where the power spectrum deviates from the best-fit Planck value 
\cite{Planck2015I}.

We conclude that there are three main features found in our investigation: 
the SEVEM map is affected by the Galactic Pole and Galactic Center directions and when used for full-sky analyses a careful treatment of its processing algorithm should be taken into account. The other three maps agree remarkably well, except with respect to 
their alignment towards the Galactic Center. The cosmic dipole is the only considered 
physical direction, for which we are able to identify an effect on all full-sky maps. 
The alignments are localized in multipole space and stem from three ranges 
$l \in [2,5]$, $l$ around $20$ and $l$ around $42$.

\subsection{Comparison of directions using likelihood histograms}

By plotting histograms for the likelihoods introduced in Eqs.~\eqref{Louter} and \eqref{Linner} on logarithmic intervals $[0,1)$, $[1,10)$ and $[10,100]$ for the real CMB full-sky data and comparing them to the expectation from Gaussian and isotropic Monte Carlo simulations, we obtain a measure of anisotropy on the whole range $2 \leq l \leq 50$. The results confirm the findings of the last section. Furthermore, combining two statistics into one likelihood compresses the information content.

\subsubsection{SEVEM}

\begin{figure}
\includegraphics[width = \hsize]{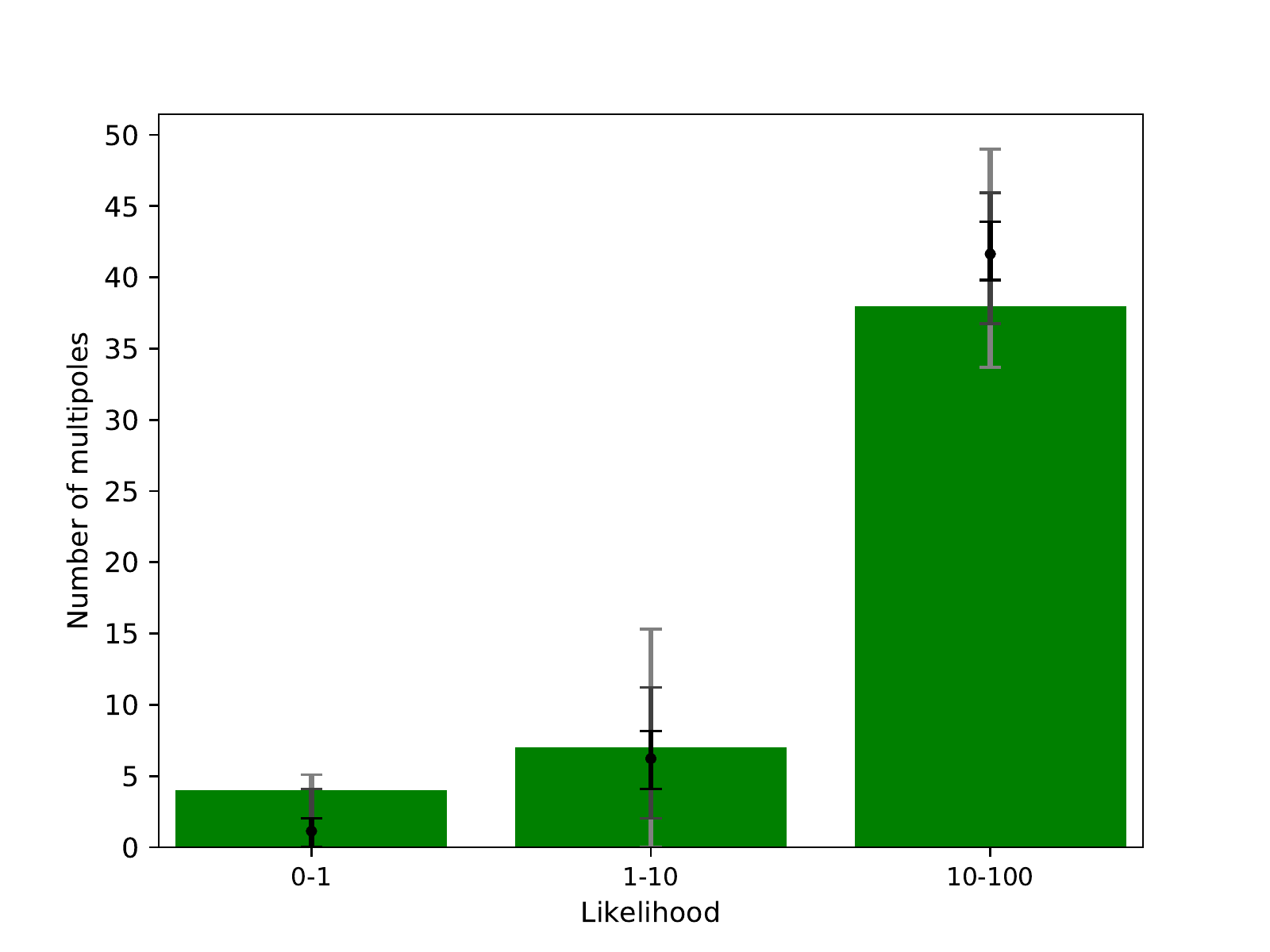}
\caption{Inner likelihood histogram for SEVEM. Gaussian, isotropic expectation and $1/2/3\sigma$ regions (black, gray and lightgray, respectively) are included.}
\label{SEVEM_inner}
\end{figure}

\begin{figure}
\includegraphics[width = \hsize]{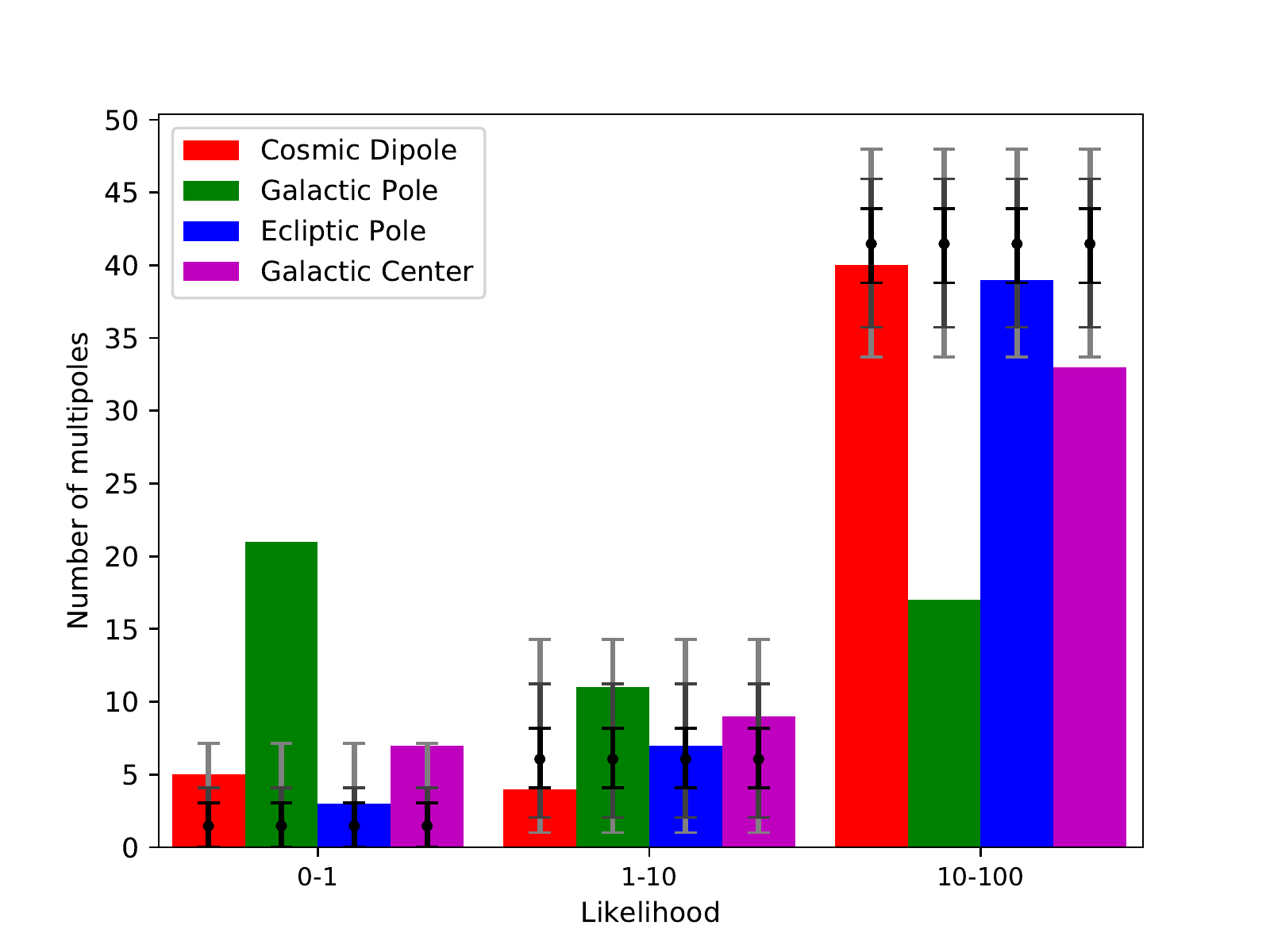}
\caption{Outer likelihood histogram for SEVEM. Gaussian, isotropic expectation and $1/2/3\sigma$ regions (black, gray and lightgray, respectively) are included.}
\label{SEVEM_outer}
\end{figure}

The large deviations from the expectation, which have been observed by investigating the alignment statistics alone, can be seen more easily from the likelihood histograms of the inner likelihood (see Fig.~\ref{SEVEM_inner}) and the outer likelihood (see Fig.~\ref{SEVEM_outer}).

The bin $[0,1)$ in Fig.~\ref{SEVEM_inner} contains too many multipoles at the $2\sigma$ level, while the bin of largest multipoles $[10,100]$ contains too few multipoles between the $1$ and $2\sigma$ level. This shows that likelihoods are shifted from the largest to the smallest values.

SEVEM's strange behavior becomes even more pronounced when considering the outer likelihoods, see Fig.~\ref{SEVEM_outer}. While the ecliptic pole and the cosmic dipole show anomalous behavior at the $2\sigma$ level, the Galactic Center ($3\sigma$) and especially the Galactic Pole ($\gg 3\sigma$) are far off from the expectation. Altogether, 21 out of 49 values for the Galactic Pole have a likelihood which is smaller than $1 \%$, which is far beyond $3\sigma$.

Hence, we can conclude that SEVEM shows a slight anomaly with respect to intramultipole correlations, while it shows an enormous anomaly with respect to outer correlations with the Galactic Center and most strongly with the Galactic Pole, whose statistics correspond to measures of the influence of the Galactic Plane. The combination of the Galactic Pole and Center anomalies evokes the conjecture, that SEVEM is influenced by the Milky Way when no masking procedure is considered.

\subsubsection{NILC}

\begin{figure}
\includegraphics[width = \hsize]{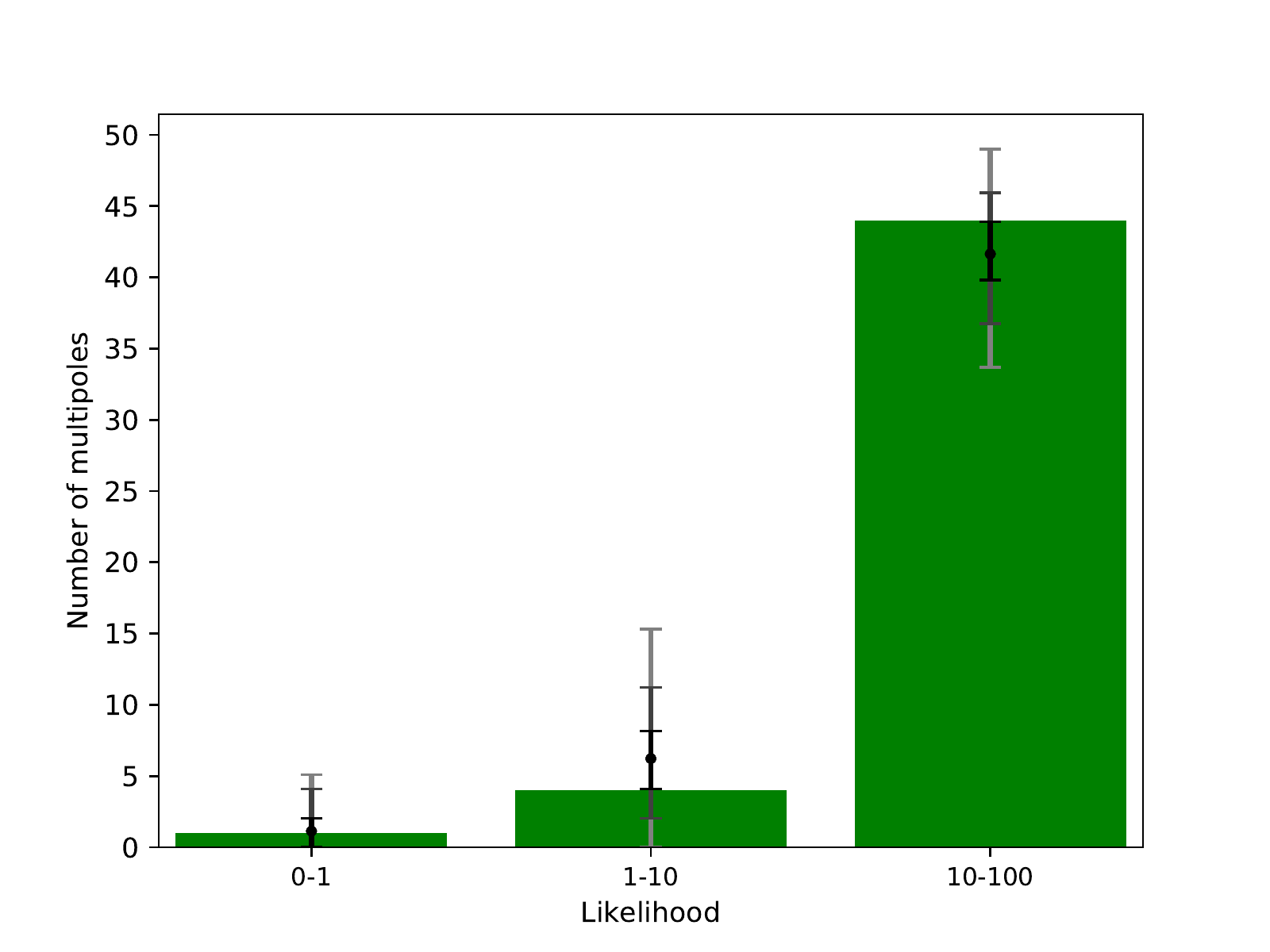}
\caption{Inner likelihood histogram for NILC. Gaussian, isotropic expectation and $1/2/3\sigma$ regions (black, gray and lightgray, respectively) are included.}
\label{NILC_inner}
\end{figure}

\begin{figure}
\includegraphics[width = \hsize]{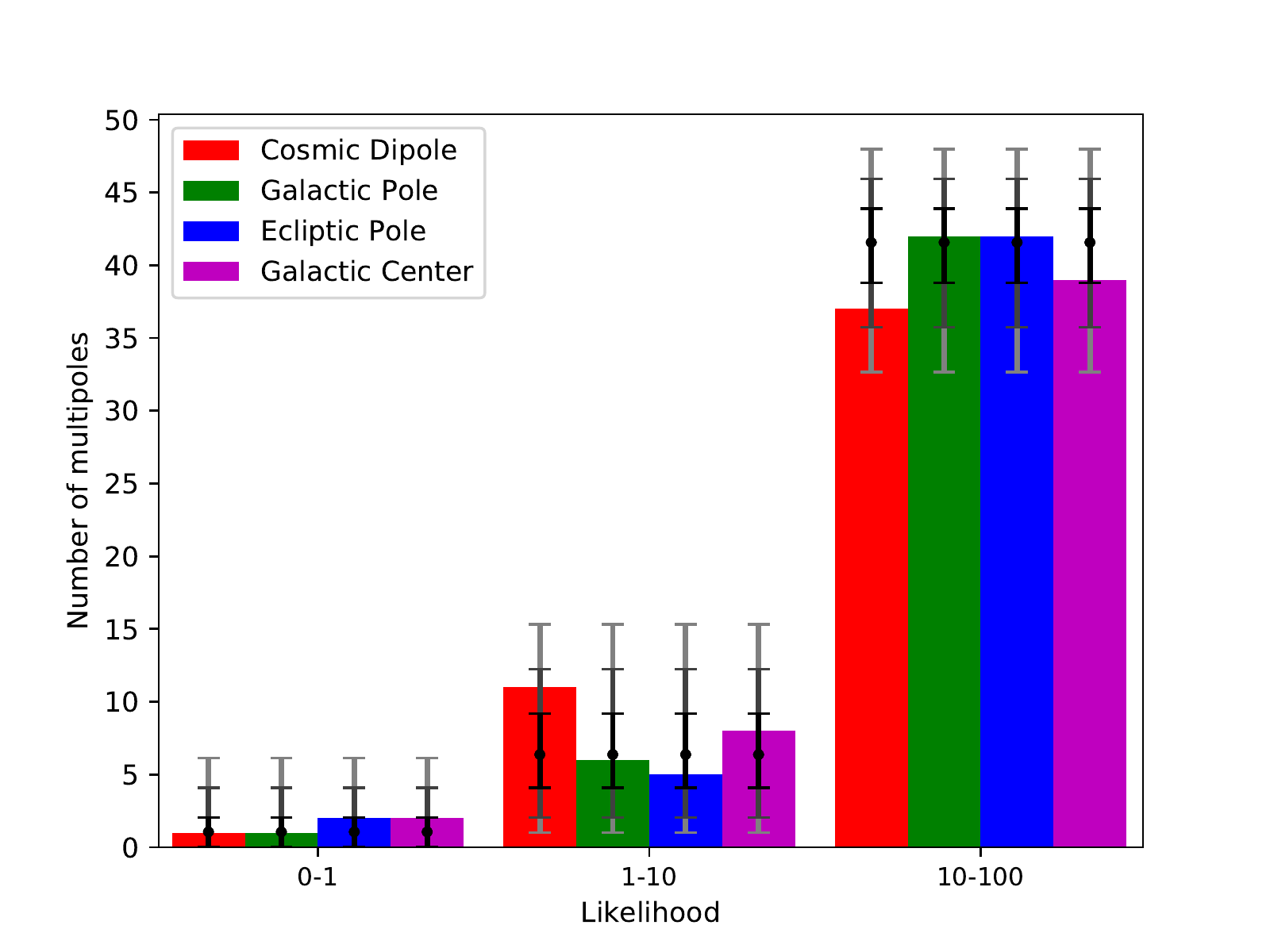}
\caption{Aligned likelihood histogram for NILC. Gaussian, isotropic expectation and $1/2/3\sigma$ regions (black, gray and lightgray, respectively) are included.}
\label{NILC_aligned}
\end{figure}

\begin{figure}
\includegraphics[width = \hsize]{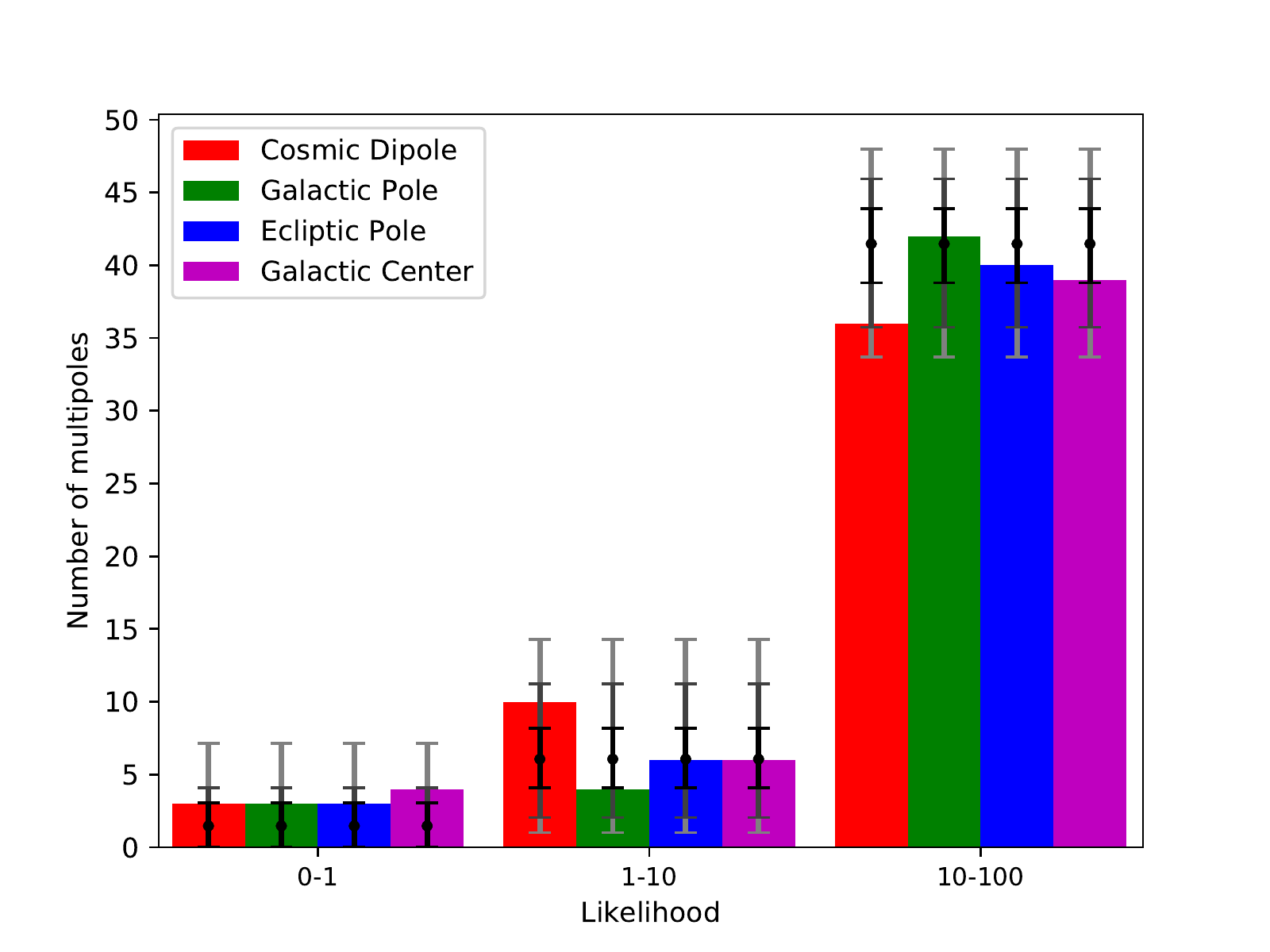}
\caption{Outer likelihood histogram for NILC. Gaussian, isotropic expectation and $1/2/3\sigma$regions (black, gray and lightgray, respectively) are included.}
\label{NILC_outer}
\end{figure}

\begin{figure}
\resizebox{\hsize}{!}{\includegraphics{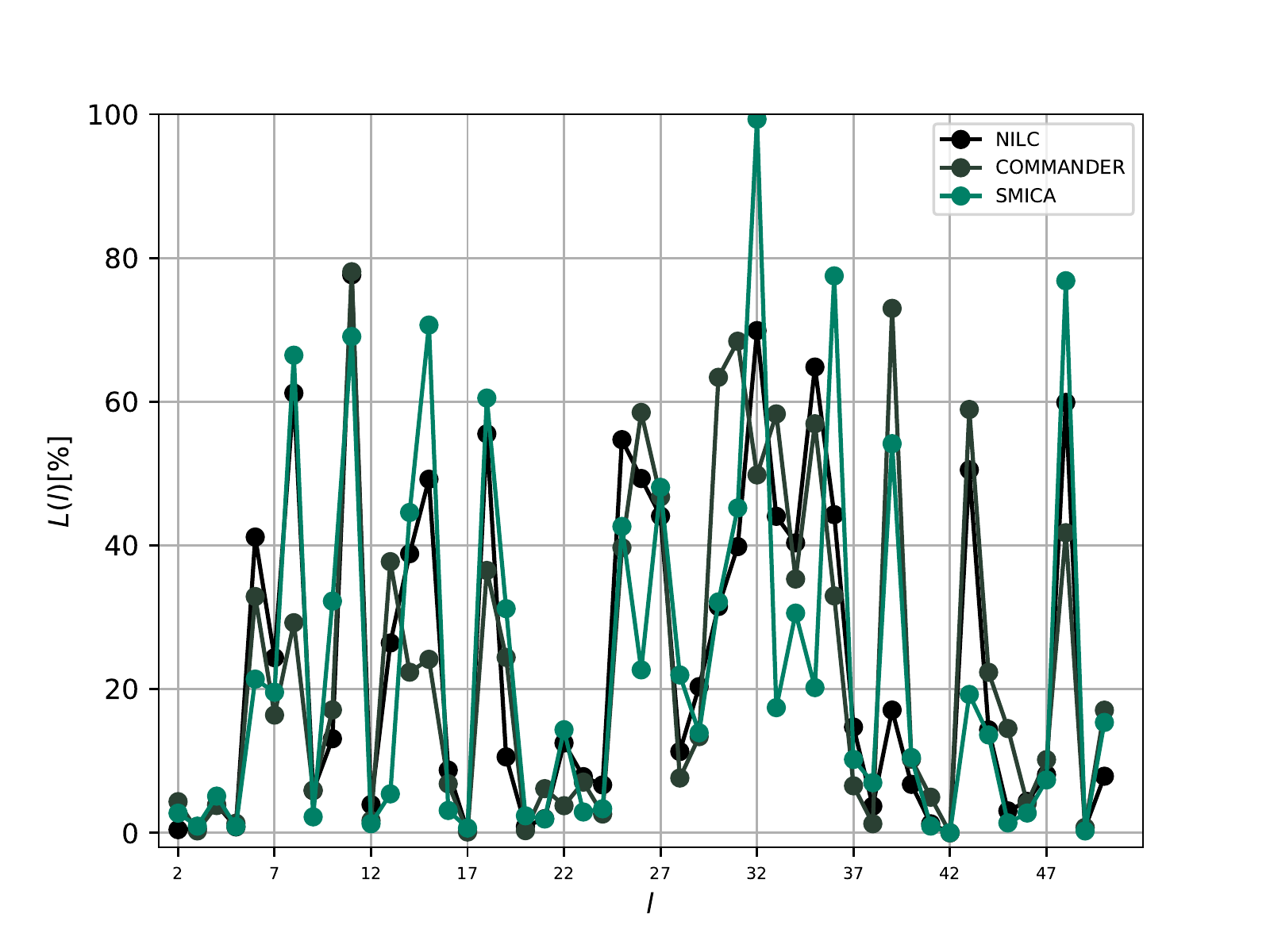}}
\caption{Comparison of outer likelihoods in dependence of $l$ for the range $2 \leq l \leq 50$ and the cosmic dipole as the outer direction. SEVEM has been excluded.}
\label{LH_outer_dip}
\end{figure}

Now, we exclude SEVEM and only consider the other three maps: COMMANDER, NILC and SMICA. It turns out, as already conjectured in the investigation of the pure statistics, that all three maps deviate only marginally. While COMMANDER tends to be the map with slightly larger likelihoods than the other two maps, NILC is equipped with the smallest confidence mask and therefore we choose to present only the NILC results and mention that COMMANDER seems to be closer to the expectation while SMICA is slightly further away than NILC. Here again the striking similarity of all three maps, despite their very different cleaning procedures, is quite remarkable. For (nonlogarithmic) likelihood histograms of the other maps we refer to \cite{Pinkwart}.

The overall structure of inner likelihoods (see Fig.~\ref{NILC_inner}) is remarkably normal except for the fact that the number of multipoles in the likelihood bin $[1,10)$ is too low at the $1\sigma$ level while the single multipole in the lowest bin equals the expectation value. It seems that some artificial intramultipole isotropy could have been induced in the course of data processing, resulting in a lack of variance in intramultipole correlations. Apart from this, the inner likelihoods do not show any further noticeable feature. 

The aligned likelihood, see Fig.~\ref{NILC_aligned}, shows expected behavior for all directions but the cosmic dipole. The number of multipoles with aligned likelihoods in the bin $[1,10)$ for the cosmic dipole as outer direction is higher than one would expect from a Gaussian and isotropic set of $a_{lm}$ at approx.~$1.5\sigma$.

The outer likelihood (see Fig.~\ref{NILC_outer}) shows the same behavior as the aligned likelihood but with an additional excess of likelihoods in the $[0,1)$-bin at $1\sigma$ level for the cosmic dipole, the Galactic Pole and the ecliptic pole, and at $2\sigma$ level for the Galactic Center. Considering both bins of small likelihoods $[0,1)$ and $[1,10)$ together, the cosmic dipole sticks out most again. The similar behavior of the outer and aligned likelihoods confirms the robustness of the likelihood definition against correlations of the included statistics.

Figure \ref{LH_outer_dip} shows the outer likelihoods for the cosmic dipole. It is clearly shown that there is a range, $25 \leq l \leq 34$, that does not include any unlikely data point regarding the cosmic dipole. Hence, using the method applied here, we cannot identify any statistically significant effect of the cosmic dipole on the data on angular scales of about $5.3$ 
to $7.2 \deg$. But once again one sees that in the ranges $2 \leq l \leq 5$ and
$20 \leq l \leq 24$ low likelihoods cluster. Comparing the two statistics $S_{\mathbf{D}}^v$ and $S_{\mathbf{D}}^{||}$ one sees that the contribution to low likelihoods in this range mainly stems from the vertical statistic. We conclude that the slight excess of low likelihoods regarding the cosmic dipole in the range $2\leq l \leq 50$ mainly stems from the two regions $2\leq l \leq 5$ and $20 \leq l \leq 24$.

\section{Discussion}
\label{discussion}

We find that SEVEM strongly deviates from COMMANDER, NILC and SMICA in every regard. It is strongly aligned with the Galactic Center and especially the Galactic Pole and antialigned with the dipole and the ecliptic pole in the $S^{||}_{\mathbf{D}}$ statistic. The alignment with the Galactic Center is most prominent on midrange multipoles, indicating a residual effect of the Galactic Core that has not been removed in the cleaning process. Furthermore, the deviation of SEVEM from the other maps is most present at $l \geq 12$, which indicates that the central part of the Milky Way is the dominating source of distraction in SEVEM. Nevertheless, the correlation with the cosmic dipole, that is present in the other maps, could also be seen in SEVEM. It is just overshadowed by the sizable galactic residuals. Since none of the above observations surprised, SEVEM serves us as a control map. The fact that we were able to identify the expected residual foreground features of SEVEM with our method shows that our method yields geometrically easily interpretable results and that the heuristic geometric intuition of the used statistics is correct. Hence, we propose that the outer statistics truly measure the influence of the given directions that are included.

COMMANDER, NILC and SMICA show a similar behavior, deviating only marginally on the observed range. While with respect to the Galactic Center and Pole and the ecliptic pole the data do not show abnormal statistical behavior, a correlation with the cosmic dipole is visible, concentrating mainly on largest angular scales $2 \leq l \leq 5$ and on intermediate scales $l=20,21,22,23,24$. The behavior of the aligned statistic shows an analogue scheme at both ranges; first there is antialignment, and then alignment. 

The correlation with the cosmic dipole at the lowest multipoles is present both in antialignment ($l=2,3,5$) and alignment ($l=4$), but also the range $35 \leq l \leq 45$ seems slightly conspicuous, while the range $25 \leq l \leq 34$ is surprisingly normal with the absence of small likelihoods. The large scale (anti)alignments might imply that we do not yet fully understand the true nature of the dipole, i.e. the relative motion of the Solar System with respect to the cosmic frame. Since these anomalies are present in all of the maps such a physical origin could be more likely than data processing reasons.

It should be noted that the SMICA algorithm assumes isotropy and Gaussianity from the beginning and thus it is biased. Furthermore some assumptions on the spectrum on synchrotron and free-free emission in the physics-based cleaning process of COMMANDER as well as its noise model might induce a bias on isotropy. In a weak sense also NILC might be biased. SEVEM is the only map where an \textit{a priori} bias on isotropy and Gaussianity can be excluded. This fact might influence the interpretation of the results.

\section{Conclusion}
\label{conclusion}

The purpose of this work was to study the complete randomness of the microwave sky by means of multipole vectors (MPV) in  the hope of identifying deficits in our understanding or the data analysis of CMB full-sky maps.

We gave an overview over different extraction methods for MPVs and their statistical properties. MPVs can be represented via a symmetric and trace-free tensor applied to the symmetric and trace-free product of unit vectors, which yields an algorithm for extracting MPVs from the spherical harmonic decomposition. Alternatively methods from algebraic geometry can be used to identify MPVs as lines in $\mathbb{C}\mathrm{P}^2$. A third approach uses the extension of the Bloch sphere to higher spin and the stereographic projection. The resulting polynomial can be understood as the scalar product of a spin state with Bloch coherent states. The latter approach can be used to assign joint probability densities to MPVs. It turns out that the explicit expression for the joint probability density is the same for the set of all distributions of completely random $a_{lm}$. This set forms a subset of statistical isotropic distributions and the intersection of completely random and Gaussian distributions yields the regime of standard cosmology. 

Using different simple statistics we observed numerically a correlation of the full-sky cleaned maps with the cosmic dipole on the largest angular scales $2 \leq l \leq 5$ and intermediate angular scales $l=20,21,22,23,24$. Furthermore around $l=40$ low likelihoods cluster and the multinomial p-value drops. These are the same multipole numbers which also deviate from the theoretical expectation in the angular power spectrum \cite{Planck2015I}. 
To the authors' knowledge, this "conspiracy" of MPV and power spectrum has not been observed before. Other covariances of CMB anomalies have recently been investigated in \cite{Huterer2018}.

One main conclusion we draw is that the SEVEM map is still strongly correlated with the Galactic Center and especially the Galactic Pole in our analysis. The cross-talk between MPVs and masked skies needs to be studied in more detail before one can use the foreground cleaned maps with small galactic masks for MPV analysis.

In the future, one could also study cross-multipole correlations on the observed range of scales and investigate if the previously observed large scale correlations continues down to smaller scales. 

Furthermore one needs more insight about possible physical reasons for CMB anomalies. One should especially focus on detailed studies of the dipole and reveal its true nature. Analyses of the radio sky with galaxy surveys
hint towards an increased radio dipole amplitude \cite{BlakeWall,Singal2011,RubartSchwarz2013,Tiwari2015}, 
which could be caused by an intrinsic, nonkinematic CMB dipole.

\begin{acknowledgments}
We acknowledge financial support by the DFG Research Training Group 1620 Models of Gravity. Furthermore we would like to thank Craig Copi for the multipole vector calculation program \cite{CopiProg} that we used, and Gernot Akemann for useful discussions and an insight about random matrix theory.

This work is based on observations obtained with Planck (http://www.esa.int/Planck), an ESA science mission with instruments and contributions directly funded by ESA Member States, NASA, and Canada.

The results in this paper have been derived using the HEALPix \cite{HealpixPaper} package and especially the HEALPy implementation for \textsc{python}.
 \end{acknowledgments}

\begin{appendix} 

\section{Derivation of one-point density}
\label{one-point}

For $l=1$ we get from \eqref{PDF 2l}

\begin{align*}
p_1\left(\zeta,\frac{-1}{\zeta^*}\right) &= \frac{1}{\pi}\frac{1}{|\zeta|^2}\frac{\left|\zeta + \frac{1}{\zeta^*}\right|}{\left( (1+|\zeta|^2)(1+|1/\zeta^*|^2) \right)^{3/2}} \\
&= \frac{1}{\pi}\frac{1}{(1+|\zeta|^2)^2}.
\end{align*}
Using $\zeta = \tan(\theta/2)\exp(i\phi)$ we receive
\begin{align*}
\left| \left( \frac{\partial (\zeta,\zeta^*)}{\partial (\theta,\phi)} \right) \right| &= \tan(\theta/2)(1+\tan^2(\theta/2)) \\
\Rightarrow d\Omega &= \frac{2 \sin(\theta/2)\cos(\theta/2)}{\tan(\theta/2)} \frac{d\zeta d\zeta^*}{1+|\zeta|^2} \\
&=  \frac{2d\zeta d\zeta^*}{(1+|\zeta|^2)^2},
\end{align*}
where $d\Omega = \sin(\theta)d\theta d\phi$ denotes the solid angle element and $\left| \left( \frac{\partial (\zeta,\zeta^*)}{\partial (\theta,\phi)} \right) \right|$ the Jacobi determinant of the change of coordinates. Hence we have
\begin{align*}
p_1(\theta,\phi) \frac{2d\zeta d\zeta^*}{(1+|\zeta|^2)^2} &\stackrel{!}{=} p_1(\zeta,-1/\zeta^*)d\zeta d\zeta^*  \\
\Rightarrow p_1(\theta,\phi) &= \frac{1}{2\pi}.
\end{align*}

\section{Joint Probability Distribution of Multipole Vectors: Connection to Gaussian analytic functions and random matrix theory}
\label{RMT}
The Majorana polynomial in \eqref{Majorana function} is a special case of the wide class of Gaussian analytic functions (GAFs), see \cite{Krishnapur}. In general, a GAF is defined as a random field on $\mathbb{C}^n$ such that for each $z_1,\ldots,z_n$ the quantity $f(z_1,\ldots,z_n)$ is a normally distributed random variable.

For every $L\in \mathbb{N}$ the function 
\begin{equation}
f(z) = \sum_{n=0}^L \sqrt{\binom{L}{L-n}} a_n z^n,
\end{equation}
with identically and independently distributed zero mean and unit variance complex random variables $a_{n}$ is a GAF whose zero set is invariant under the action of SO(3). Its covariance kernel is given by $\mathrm{Cov}(f(z),f(w)) = K(z,w) =  (1+zw^*)^{L}$. The Majorana polynomial equals this GAF up to a factor $(-1)^l$ which can be combined into $\Psi_m$, and with $L=2l$ and substituting $n=m+l$, yielding $ \Psi_m = a_{m+l}$. The $\Psi_m$ do not have unit variance, but variance $C_l$. By rescaling $\Psi_m$ a common factor for all $\Psi_m$ can be pulled out of the sum. This does not change the behavior of the zeros.

The general density \eqref{PDF general} holds for every GAF, while the one-point density \eqref{p1} can be expressed as 
\begin{equation}
p_1(z) = \Delta \log(K(z,z))/4\pi,
\end{equation}
for a general GAF. This equals -- up to a different normalization -- the one-point density of the Majorana polynomial in $\mathbb{C}$ which was used in the proof of \eqref{p1}. The formula above is known as the Edelman-Kostlan formula; see \cite{Edelman}.

One can show that one-point statistics, which are compactly supported, are asymptotically normal regarding rotationally invariant GAFs. Let $\phi \in \mathcal{C}^2_{c}(\Lambda)$ and 
\begin{equation}
\mathcal{L}_L(\phi) := \sum_{z \in f^{-1}(0)} \phi(z),
\end{equation}
then the following asymptotic behavior is valid:
\begin{equation}
\sqrt{L}(\mathcal{L}_L(\phi) - \langle \mathcal{L}_L(\phi)\rangle) \stackrel{l \rightarrow \infty, \, \text{distribution}}{\rightarrow} N(0,\kappa(\phi)),
\end{equation}
where $\kappa(\phi)$ denotes some number that depends on the function $\phi$. Unfortunately, the above is \textit{a priori} not true for functions $\phi$ with arbitrary support. Hence, it does not apply to the statistics in Sec.~\ref{Analysis}. Since we are dealing with one hemisphere, one could restrict the scalar products appearing in those statistics to the unit disc. This cutoff compactifies the statistic but unfortunately it destroys any kind of differentiability. Nevertheless the result above could be used to study local statistics on certain patches on the sky in the large $l$ limit in future investigations.

Remember that the Majorana polynomial has covariance kernel $K(z,w) = (1+zw^*)^{2l}$. The following statement will show that MPVs as zeros of the isotropic GAF and eigenvectors of Gaussian random matrices are tightly connected:
let $A$, $B$ be independent $(n\times n)$ random matrices with identically and independently distributed complex standard Gaussian entries. Then the eigenvalues of $A^{-1}B$ form a determinantal point process on $\mathbb{C}$ with covariance kernel $K(z,w) = (1+zw^*)^{n-1}$ with respect to the measure $n /(\pi(1+|z|^2)^{n+1}) \cdot \mathrm{d}m(z)$ and the eigenvectors are distributed as
\begin{equation}
p(\{z_i\}) = \frac{1}{n!}\left( \frac{n}{\pi} \right)^n \prod_{k=1}^n \frac{\prod_{i<j} |z_i-z_j|^2}{(1+|z_k|^2)^{(n+1)}}
\end{equation}
according to the Lebesgue measure on $\mathbb{C}^n$.

One can see that the covariance kernel of these eigenvalues and the covariance kernel of the Majorana polynomial are equal for $n=2l+1$ and that the probability density above and the one in \eqref{PDF 2l} look similar, but still different. The reason for this difference is of course that the zeros of the spherical GAF do not follow a determinantal process, but rather some kind of permanental process. In fact, the only case of a GAF whose zero set is known to follow a determinantal process is the following one:
\begin{equation}
f(z) = \sum_{n=0}^{\infty} a_n z^n.
\end{equation}
This is a special case of a hyperbolically invariant GAF and there are some striking results considering this special function. Unfortunately, the rotationally invariant GAF has not yet been confirmed to imply a determinantal process.

A better understanding of the possible relationship between Gaussian random matrices and MPVs would help in investigating CMB anomalies with MPVs. Two-point functions of eigenvalues are known simple expressions while in principle the joint probability density of the MPVs \eqref{PDF general} is computable as well.

\section{Stereographic Projection}
\label{app_stereo}

\begin{figure}[!ht]
\begin{subfigure}[c]{\hsize}
\includegraphics[width = \hsize]{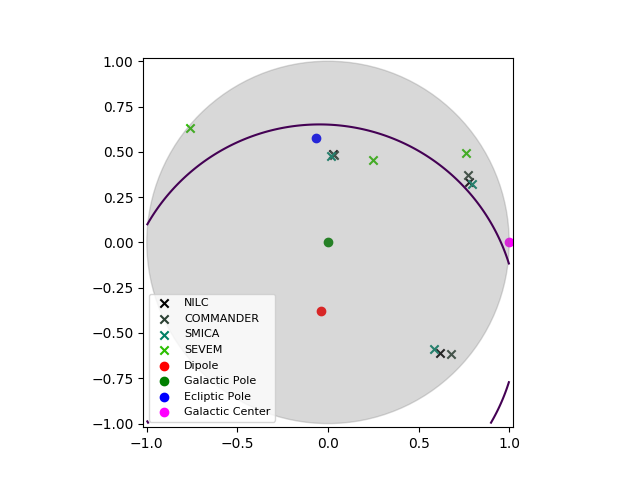}
\subcaption{$l=3$}
\label{stereo_3}
\end{subfigure}
\begin{subfigure}[c]{\hsize}
\includegraphics[width = \hsize]{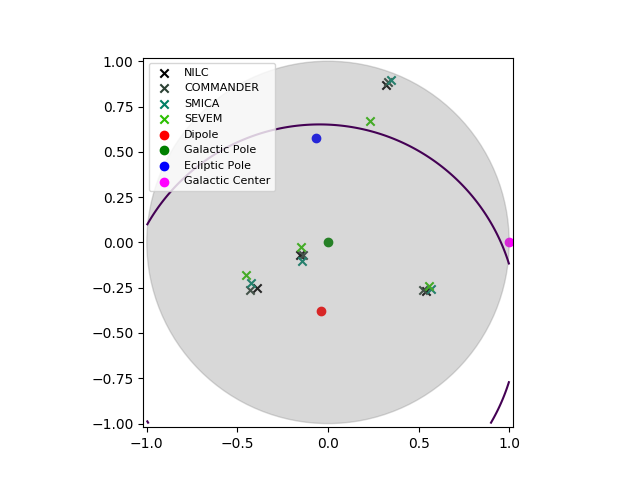}
\subcaption{$l=4$}
\label{stereo4}
\end{subfigure}
\caption{Multipole vectors and physical directions in stereographic projection. The violet curve shows the plane orthogonal to the cosmic dipole.}
\label{stereo_low}
\end{figure}

\begin{figure}[!ht]
\begin{subfigure}[c]{\hsize}
\includegraphics[width = \hsize]{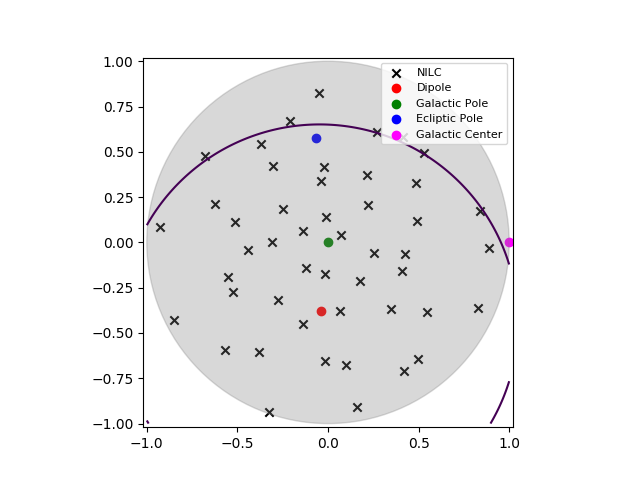}
\subcaption{$l=48$}
\label{stereo_48}
\end{subfigure}
\begin{subfigure}[c]{\hsize}
\includegraphics[width = \hsize]{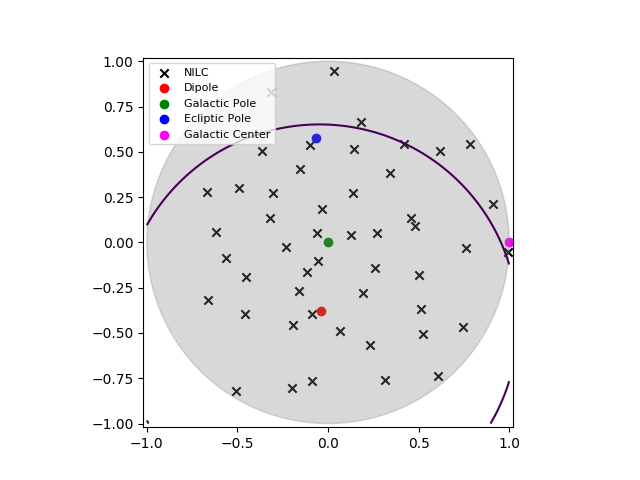}
\subcaption{$l=49$}
\label{stereo_49}
\end{subfigure}
\caption{Multipole vectors (only NILC) and physical directions in stereographic projection. The violet curve shows the plane orthogonal to the cosmic dipole.}
\label{stereo_high}
\end{figure}

Here we show the stereographical projection of multipole vectors for all full-sky cleaned maps together with the stereographic projection of four physical directions for $l=3,4,48,49$. We choose to show the small multipoles $l=3,4$ in Fig.~\ref{stereo_low} because the statistics in the main text show that in all maps the quadrupole is unusually weakly aligned with the cosmic dipole and $l=4$ is unusually strongly aligned with the cosmic dipole. We also choose to plot the stereographic projection for two higher values of $l$ in Fig.~\ref{stereo_high}, one of which ($l=48$) is close to the expectation regarding alignment with the cosmic dipole and one of which ($l=49$) is especially weakly aligned with the cosmic dipole.

\section{Alignment statistics for masked maps}
\label{app_mask}

\begin{figure}[!ht]
\begin{subfigure}[c]{\hsize}
\includegraphics[width = \hsize]{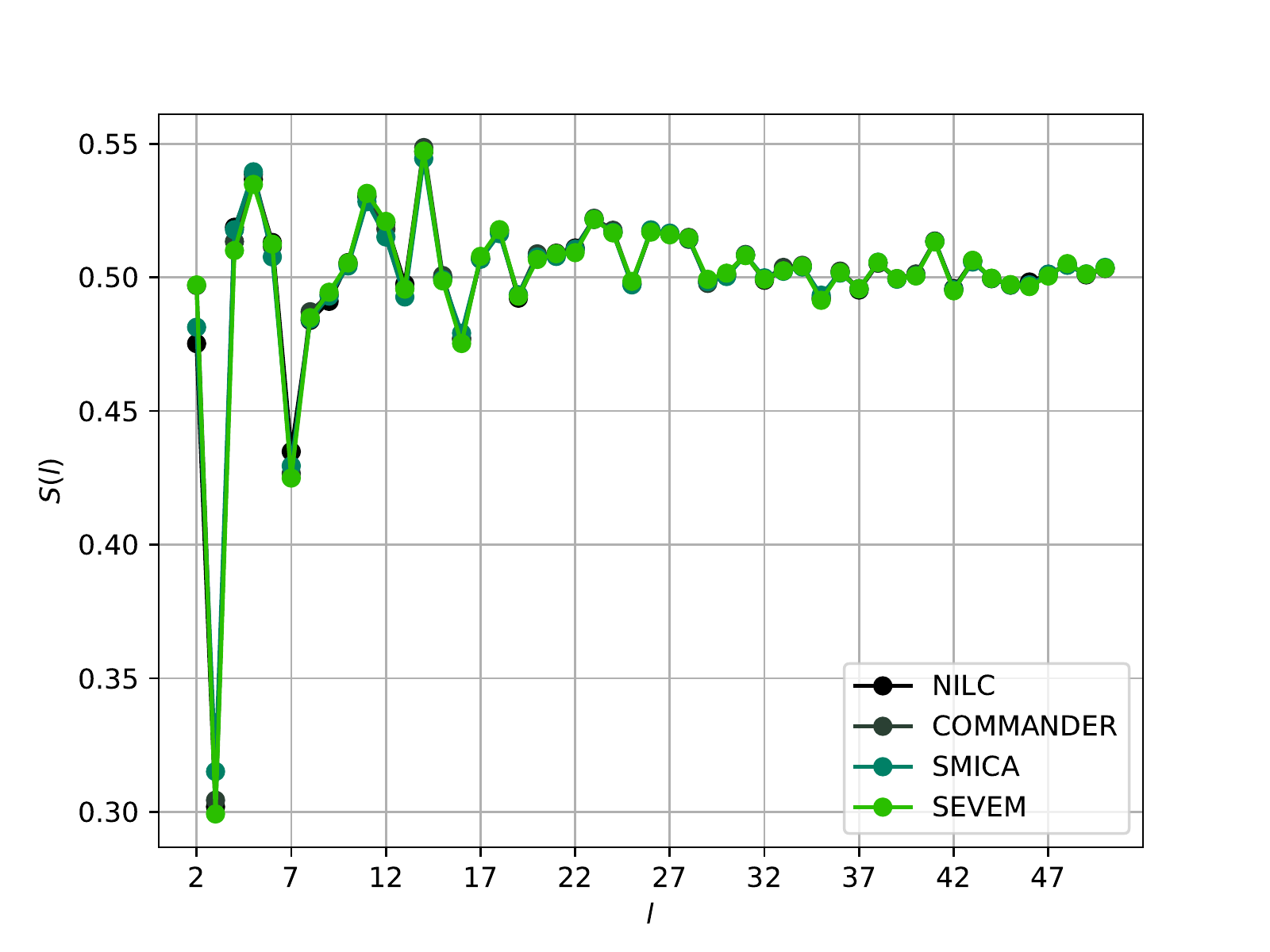}
\subcaption{$S^{||}_{\mathbf{D}}$ with $\mathbf{D}$ the Galactic Pole}
\label{With_mask_gp}
\end{subfigure}
\begin{subfigure}[c]{\hsize}
\includegraphics[width = \hsize]{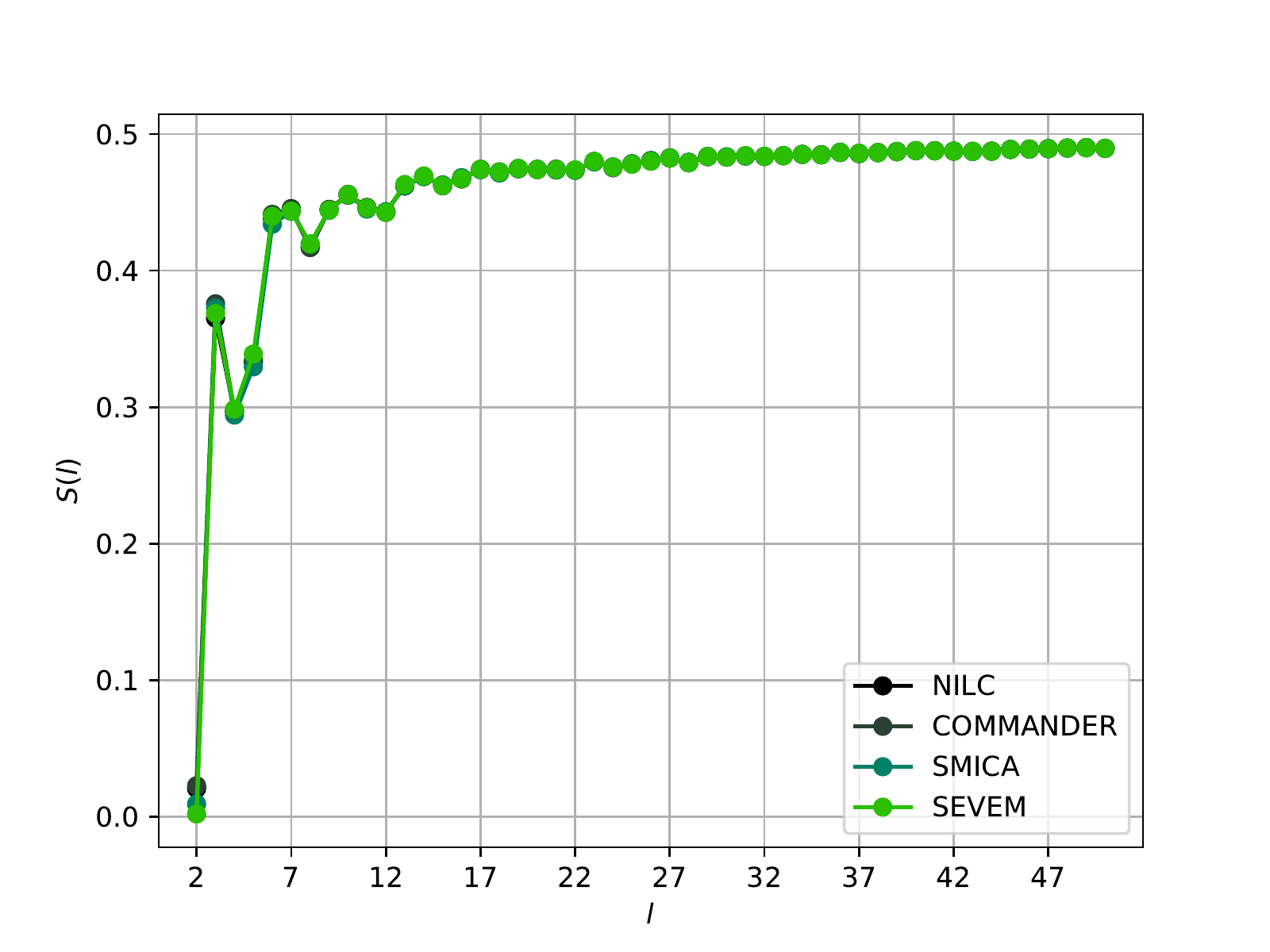}
\subcaption{$S^{||}$}
\label{With_mask_aligned}
\end{subfigure}
\caption{Aligned statistics for the Planck data with the SEVEM mask applied and without Monte Carlos from $l=2$ to $l=50$. Once the Galactic Center is properly masked, all four 
foreground cleaned maps agree very well with each other. The strong deviation of the aligned statistics from the generic expectation for small $l$ is due to the mask, which 
is the reason why masked maps cannot be used for the analysis of statistical isotropy of the lowest multipole moments.} 
\label{With_mask}
\end{figure}

In Fig.~\ref{With_mask} we show the aligned statistics for all maps after the SEVEM confidence mask has been applied to all four maps collectively.

\end{appendix}
\clearpage
\bibliography{./mybiblio.bib}

\end{document}